\documentclass[11pt]{article}

\usepackage{amsmath}
\usepackage{bm}
\usepackage{amssymb}
\usepackage{amstext}
\usepackage{latexsym}
\usepackage{color}
\usepackage{graphicx}
\usepackage{epsfig}
\usepackage{subfigure}
\usepackage{rotating}
\usepackage{curves}
\usepackage{epic}
\usepackage{amsthm}
\allowdisplaybreaks[1]

\usepackage{amsfonts}
\usepackage[T1]{fontenc}
\usepackage[latin1]{inputenc}
\usepackage{authblk}
\usepackage{tikz}
\usetikzlibrary{shapes,arrows}
\usepackage{graphics}
\usepackage{verbatim}
\usepackage{color}
\usepackage{hyperref}
\usepackage{mathtools}

\newcommand{\be}{\begin{equation}}
\newcommand{\en}{\end{equation}}

\newtheorem{thm}{Theorem}
\newtheorem{cor}[thm]{Corollary}
\newtheorem{prop}[thm]{Proposition}

\newtheorem{proposition}[thm]{Proposition}

\newtheorem{defi}{Definition}[section]
\newtheorem{lem}[defi]{Lemma}

\newtheorem{Theo}{Theorem}[section]

\newtheorem{remark}[Theo]{Remark}

\newcommand{\bedefin}{\begin{defi}}
\newcommand{\findefi}{\end{defi} \medskip}

\newcommand{\betheo}{\begin{theorem}$\!\!${\bf \,\,\,}}
\newcommand{\entheo}{\end{theorem}}
\newcommand{\enth}{\end{theorem}}

\newcommand{\becor}{\begin{cor}$\!\!${\bf .}}
\newcommand{\encor}{\end{cor}}

\newcommand{\belem}{\begin{lem}$\!\!${\bf }}
\newcommand{\enlem}{\end{lem}}

\newcommand{\prf}{\noindent{\bf{ Proof}\,\,}}

\newcommand{\bea}{\begin{eqnarray}}
\newcommand{\ena}{\end{eqnarray}}

\newcommand{\beano}{\begin{eqnarray*}}
\newcommand{\enano}{\end{eqnarray*}}

\newcommand{\bee}{\begin{enumerate}}
\newcommand{\ene}{\end{enumerate}}

\newcommand{\bei}{\begin{itemize}}
\newcommand{\eni}{\end{itemize}}

\newcommand{\betab}{\begin{tabular}}
\newcommand{\entab}{\end{tabular}}

\newcommand{\bd}{\begin{displaymath}}

\newcommand{\h}{{\mathfrak H}}

\newcommand{\nn}{\nonumber}

\newcommand{\ba}{\mathbf a}
\newcommand{\bb}{\mathbf b}

\newcommand{\bp}{\mathbf p}
\newcommand{\bq}{\mathbf q}
\newcommand{\br}{\mathbf r}
\newcommand{\bs}{\mathbf s}

\newcommand{\bx}{\mathbf x}

\newcommand{\g}{G_{\hbox{\tiny{NC}}}}
\newcommand{\gd}{\widehat{G}_{\hbox{\tiny{NC}}}}
\newcommand{\G}{\mathfrak{g}_{\hbox{\tiny{NC}}}}
\DeclareMathOperator{\tr}{Tr}

\catcode `\@=11 \@addtoreset{equation}{section}

\catcode `\@=12

\begin{document}

\title{Wigner Functions for Noncommutative Quantum Mechanics:\\
a group representation based construction}
\author[1,2]{S. Hasibul Hassan Chowdhury\thanks{shhchowdhury@gmail.com}}
\author[2]{S. Twareque Ali\thanks{twareque.ali@concordia.ca}}
\affil[1]{Chern Institute of Mathematics, Nankai University, Tianjin 300071, P. R. China}
\affil[2]{Department of Mathematics and Statistics, Concordia University, Montr\'eal, Qu\'ebec, Canada H3G 1M8}
\date{\today}

\maketitle

\begin{abstract}
This paper is devoted to the construction and analysis of the Wigner functions for noncommutative quantum mechanics, their marginal distributions and star-products, following a technique developed earlier, {\it viz\/,}  using the unitary irreducible representations of the group $\g$, which is the three fold central extension of the abelian group of $\mathbb R^4$. These representations have been exhaustively studied in earlier papers. The group $\g$ is identified with the kinematical symmetry group of noncommutative quantum mechanics of a system with two degrees of freedom. The Wigner functions studied here reflect different levels of non-commutativity -- both the operators of position and those of momentum not commuting, the position operators not commuting and finally, the case of standard quantum mechanics, obeying the canonical commutation relations only.
\end{abstract}

\section{Introduction}\label{sec:intro}
This paper deals with the problem of the definition and construction of Wigner functions for a model of noncommutative quantum mechanics (NCQM) with two degrees of freedom. In earlier papers \cite{ncqmjmp,ncqmjpa} we have used a group theoretical analysis to obtain the structure of such a quantum system. In the first instance, we used the unitary irreducible representations of a doubly centrally extended version of the Galilei group, in two space and one time dimensions, to obtain the commutation relations of NCQM, where in addition to the usual canonical commutation relations between position and momentum, the two position operators are also noncommutating. The operators of position, momentum, angular momentum and the free Hamiltonian appear as generators of these representations. The group related coherent states, arising from these representations, and labeled by the points of the phase space of the system, were then used to carry out a coherent state quantization of the phase space variables, $q$ and $p$, which also led to the correct commutation relations of noncommutative quantum mechanics. All this is in complete analogy with the well-known fact that the representations of the singly extended Galilei group are the ones defining standard quantum mechanics. A further extension of these results for this model of noncommutative quantum mechanics, where the two operators of momentum are also noncommuting, was obtained by looking at a triple central extension of $\mathbb R^4$. Consequently, this particular group was identified as the kinematical group of noncommutative quantum mechanics, in the same way as the Weyl-Heisenberg group underlies standard quantum mechanics. The Wigner function \cite{Wigner} is a well-known quantum mechanical (pseudo-) distribution function, used extensively in molecular and quantum optical computations, as well as in various theoretical contexts. A group theoretical construction of the Wigner function  was detailed in \cite{AlAtChWo,AlKrMu,plancherel}, where the standard Wigner distribution was obtained as a function on a coadjoint orbit of the Weyl-Heisenberg group. (As is well known, these are the orbits under the natural, i.e., coadjoint action of the group on the dual of its Lie algebra.) The method allowed for an extension to a large class of other groups, in particular nilpotent groups and groups which admit square integrable representations. In all these cases the Wigner function is defined on coadjoint orbits of the group (admitting  natural symplectic structures). The thrust of this paper is to define  Wigner distributions within the above framework, using the unitary irreducible representations of the triply extended group of $\mathbb R^4$. These are then taken to be the natural Wigner functions associated to noncommutative quantum mechanics.

Considering the triply centrally extended group of $\mathbb R^4$, denoted by $\g$, as the defining group  of noncommutative quantum mechanics (NCQM), we know from \cite{ncqmjpa}, that its coadjoint orbits are of three different types,  having dimensions dimensions  4, 2 and 0, respectively, all foliated inside the dual, $\G^{*}$, of the Lie algebra of $\g$. The geometry of these orbits  has been discussed in detail in \cite{ncqmjpa}. The dual of $\g$ (i.e., the set of all equivalence classes of its unitary irreducible representations), henceforth denoted by $\gd$, is characterized by 3 real parameters $\rho$, $\sigma$ and $\tau$. The elements of the dual, which  will be denoted by $U^{\rho}_{\sigma,\tau}$, are in in 1-1 correspondence with the underlying coadjoint orbits of $\g$. For nonzero values of $\rho$, $\sigma$ and $\tau$, the corresponding coadjoint orbits are either $4$-dimensional or $2$-dimensional, depending on whether the value of $\rho^{2}\alpha^{2}-\gamma\beta\sigma\tau$ is nonzero or 0, respectively ($\alpha, \beta$ and $\gamma$ being dimensional constants, whose roles will become clear later). We shall primarily focus on the $4$-dimensional coadjoint orbits $\mathcal{O}^{\rho,\sigma,\tau}_{4}$ of $\g$ with nonzero values of $\rho$, $\sigma$ and $\tau$ satisfying $\rho^{2}\alpha^{2}-\gamma\beta\sigma\tau\neq 0$ and compute the Wigner functions associated with them, using techniques involving the Plancherel inversion formula, as described in \cite{plancherel}. The Wigner function, thus obtained from the defining group $\g$ of NCQM, will be called the {\it noncommutative Wigner function} and be abbreviated as NC Wigner function in the sequel. As shown in \cite{ncqmjpa}, the unitary irreducible representations of a 2-dimensional model of  NCQM, in which the two position operators are noncommuting, but the two momentum operators commute, correspond to a family of $4$-dimensional coadjoint orbits, $\mathcal{O}^{\rho,\sigma,0}_{4}$, for nonzero values of $\rho$ and $\sigma$. We will also compute the Wigner function associated to these coadjoint orbits of $\g$. There is another 1-parameter  family of 4-dimensional coadjoint orbits, $\mathcal{O}^{\rho,0,0}_{4}$ (nonzero values of $\rho$ only), which yield the unitary irreducible representations giving standard nonrelativistic quantum mechanics. We shall compute the Wigner functions corresponding to these coadjoint orbits as well and compare our result with the original quantum mechanical Wigner function.

Following the computations of the Wigner functions associated with NCQM, we obtain the relevant marginal distributions by integrating out the position and momentum coordinates. When one carries out the integration in noncommuting position and momentum coordinates (see (\ref{int-vrbls-reprmtrzn}) below towards the end of Section \ref{subsec:wig-fcn}), one essentially ends up with the classical marginal probability distributions in the same noncommuting coordinates. But when the integration is carried out with respect to the commuting coordinates, i.e. the coadjoint orbit coordinates, one obtains marginal distributions which can be expressed suitably defined $\star$-products that are similar to those obtained by Bastos and co-workers in (\cite{bastosjmp,bastoscmp}).

Before ending this section we recall that the standard quantum mechanical cross-Wigner function \cite{Wigner} for a rank one operator $X = \vert \phi \rangle\langle \psi \vert$, with $\phi, \psi \in L^2 (\mathbb R^n, d\bx )$ has the form
\be
  \mathcal W^{\text{qm}}(X\mid \bq , \bp; h) = \frac 1{h^n} \int_{\mathbb R^n} \overline{\psi \left( \bq - \frac \bx{2}\right)}\; \exp\left[-\frac {2\pi i \bx\cdot\bp}h \right]\; \phi \left( \bq + \frac \bx{2}\right)\; d\bx , \quad \bq, \bp, \bx \in \mathbb R^n .
\label{orig-wigfcn}
\en
 This has the well-known properties of marginality, reality (when $\phi = \psi$), etc. Moreover,
 $(\bq,\bp)$ may be considered as labeling the points of an appropriate coadjoint orbit of the Weyl-Heisenberg group. The Wigner functions that we shall compute later in this paper will all have this same general form.

The rest of this paper is organized as follows. In Section  \ref{sec:wigfunc} we define the group $\g$ and write down its unitary irreducible representations that we use to construct Wigner functions. We then outline the method used to compute Wigner functions using these representations and work out a general expression for a Wigner function on a first set of orbits, with none of the three parameters $\rho, \sigma, \tau$ equal to zero. In Section (\ref{sec:wigfuncotherrep}) we construct Wigner functions for the cases where first $\tau =0$ and then where both $\sigma$ and $\tau$ equal zero. In Section \ref{sec:marginals} we compute the marginal distributions for the various Wigner functions and star-products of Wigner functions and compare them to similar results appearing in the literature. We conclude in Section \ref{sec:conclusion} with some comments about the generality of our method and point to some possible future work. Some details of computations are collected in the Appendix.

\section{NC Wigner function from the Landau gauge representation corresponding to the coadjoint orbits \texorpdfstring{$\mathcal{O}^{\rho,\sigma,\tau}_{4}$}{OGNC} of \texorpdfstring{$\g$}{GNC}}\label{sec:wigfunc}
Let us first look at the group $\g$ and its unitary irreducible representations in some detail.

\subsection{The group \texorpdfstring{$\g$}{GNCQM} and its unitary irreducible representations}\label{subsec:grp_rep}
The group $\g$ is a seven parameter, real, nilpotent group. We shall write a general element of the group as
\be
  g = (\theta, \phi, \psi, \bq , \bp), \qquad \theta, \phi, \psi \in \mathbb R, \quad \bq = (q_1, q_2) \in \mathbb R^2, \quad \bp = (p_1, p_2) \in \mathbb R^2 ,
\label{group-elem}
\en
with the group multiplication given by
\begin{eqnarray}\label{grplawtriplyextendedgrp}
\lefteqn{(\theta,\phi,\psi,\bq,\bp)(\theta^{\prime},\phi^{\prime},\psi^{\prime},
\bq^{\prime},\bp^{\prime})}\nonumber\\
&&=(\theta+\theta^{\prime}+\frac{\alpha}{2}[\langle\bq \cdot \bp^{\prime}\rangle-\langle
\bp \cdot\bq^{\prime}\rangle],\phi+\phi^{\prime}+\frac{\beta}{2}[\bp\wedge
\bp^{\prime}],\psi+\psi^{\prime}+\frac{\gamma}{2}[\bq\wedge\bq^{\prime}],
\nonumber\\
&&\;\;\;\; \bq+\bq^{\prime},\bp+\bp^{\prime}).
\end{eqnarray}
where, for two 2-vectors, $\ba = (a_1, a_2), \bb = (b_1, b_2 ), \; \ba \cdot \bb = a_1 b_1 + a_2 b_2$ and $\ba\wedge \bb = a_1 b_2 - a_2 b_1$. The quantities $\alpha, \beta$ and $\gamma$ are dimensional constants. The three parameters $\theta, \phi, \psi$ constitute the centre of the group, while the $\bq$ and $\bp$, are the parameters of $\mathbb R^4$, the three-fold central extension of which leads to $\g$. Denoting the centre by $\mathcal Z$, we see that $\g/\mathcal Z \simeq \mathbb R^4$.

We next write down the unitary irreducible representations, $U^{\rho}_{\sigma,\tau}$ of $\g$, associated to the 4-dimensional coadjoint orbits $\mathcal{O}^{\rho,\sigma,\tau}_{4}$, with all three parameters $\rho, \sigma, \tau$ non-vanishing and $\rho^{2}\alpha^{2}-\gamma\beta\sigma\tau\neq 0$ (see equation (3.4) of \cite{ncqmjpa}). These are carried by Hilbert spaces which are all copies of  $L^2 (\mathbb R^2 , dr_1 \; ds_2 )$, and have the form,
\begin{eqnarray}\label{eq:UIR-gen-NCQM}
\lefteqn{(U^{\rho}_{\sigma,\tau}(\theta,\phi,\psi,q_{1},q_{2},p_{1},p_{2})f)(r_{1},s_{2})}\nonumber\\
    &&=e^{i\rho(\theta+\alpha q_{2}s_{2}+\alpha p_{1}r_{1}+\frac{\alpha}{2}q_{1}p_{1}-\frac{\alpha}{2}q_{2}p_{2})}e^{i\sigma(\phi+\beta p_{1}s_{2}-\frac{\beta}{2}p_{1}p_{2})}\nonumber\\
    &&\times e^{i\tau(\psi+\gamma q_{2}r_{1}+\frac{\gamma}{2}q_{2}q_{1})}f(r_{1}+q_{1},s_{2}-p_{2}), \quad f \in L^2 (\mathbb R^2 , dr_1 \; ds_2 ).
\end{eqnarray}
Note that in (\ref{eq:UIR-gen-NCQM}), $r_{1}$ has the dimension of distance and $s_{2}$ has that of momentum. Taking the Fourier transform of (\ref{eq:UIR-gen-NCQM}) in the first coordinate $r_1$ yields
\begin{eqnarray}\label{eq:UIR-FT-gen-NCQM}
\lefteqn{(\hat{U}^{\rho}_{\sigma,\tau}(\theta,\phi,\psi,q_{1},q_{2},p_{1},p_{2})\hat{f})(s_{1},s_{2})}\nonumber\\
    &&=e^{-i\rho(\theta+\alpha q_{1}s_{1}+\alpha q_{2}s_{2}-\frac{\alpha}{2}q_{1}p_{1}-\frac{\alpha}{2}q_{2}p_{2})}e^{-i\sigma(\phi+\beta p_{1}s_{2}-\frac{\beta}{2}p_{1}p_{2})}\nonumber\\
    &&\times e^{-i\tau(\psi-\frac{\gamma}{2}q_{1}q_{2})}\hat{f}(s_{1}-p_{1}-\frac{\tau\gamma}{\rho\alpha}q_{2},s_{2}-p_{2}),
\end{eqnarray}
where now $\hat{f}\in L^{2}(\hat{\mathbb{R}}^{2},ds_{1}ds_{2})$. For each triple of non-zero real numbers $(\rho, \sigma, \tau)$, let $\h^\rho_{\sigma,\tau}$ denote a copy of  $L^{2}(\hat{\mathbb{R}}^{2},ds_{1}ds_{2})$. We shall compute Wigner functions for this set of representations, so that following the general construction outlined in  \cite{AlAtChWo,AlKrMu,plancherel}, they will appear as phase space functions (i.e., functions on the corresponding coadjoint orbits of $\g$), associated to Hilbert-Schmidt operators on $\h^\rho_{\sigma,\tau}$.

The  dual $\gd$ of $\g$ is naturally equipped with a measure, $d\nu_{\g}$, known as the Plancherel measure. The elements of $\gd$  are, in the present case, in one-to-one correspondence  with the coadjoint orbits \cite{Kirillovbook} in $\G^{*}$, so that $d\nu_{\g}$ can be thought of as a measure carried by the set of these orbits. From our earlier discussion, it is seen that this measure decomposes into several  disjoint parts, corresponding to the cases where none of the parameters $\rho, \sigma, \tau$ are zero and where one or more of these are zero. For the purposes of the present paper, we shall only look at the cases: (1) none of the three parameters is zero, (2) $\rho \neq 0, \; \sigma \neq 0,\; \tau = 0$  and (3) $\rho \neq 0, \; \sigma = \tau = 0$. In the first case, the set of orbits for which  $\rho^2 \alpha^2 -\gamma\beta\sigma\tau = 0$, will also not concern us here.

The Plancherel measure is usually computed using a general orthogonality relation, which can be stated in different forms, but for our purposes we shall use the form introduced in  \cite{plancherel} (see also Section \ref{subsec:wig-fcn} below):
\begin{eqnarray}\label{eq:ortho cond}
\lefteqn{\int_{\g}\left[\int_{\gd}\tr(\hat{U}^{\rho}_{\sigma,\tau}(g)^{*}A^{1}(\rho,\sigma,\tau)
C_{\rho,\sigma,\tau}^{-1})d\nu_{\g}(\rho,\sigma,\tau)\right.}\nonumber\\
&\left.\times\int_{\gd}\tr(\hat{U}^{\rho^{\prime}}_{\sigma^{\prime},\tau^{\prime}}(g)^{*}A^{2}
(\rho^{\prime},\sigma^{\prime},\tau^{\prime})C^{-1}_{\rho^{\prime},\sigma^{\prime},
\tau^{\prime}})d\nu_{\g}(\rho^{\prime},\sigma^{\prime},\tau^{\prime})\right]d\mu(g)=\langle A^{1}|A^{2}\rangle_{\mathcal{B}_{2}^{\oplus}}
\end{eqnarray}
where $\mathcal{B}^\oplus_{2}$ is the direct integral  Hilbert space
$$ \mathcal{B}^\oplus_{2} = \int^\oplus_{\gd} \mathcal B_2 (\rho, \sigma, \tau) \; d\nu_{\g}(\rho,\sigma,\tau), $$
and $\mathcal B_2 (\rho, \sigma, \tau)$ is the Hilbert space of Hilbert-Schmidt operators on $\h^\rho_{\sigma,\tau}$, with elements $A^1 (\rho,\sigma, \tau), A^2 (\rho, \sigma, \tau)$. The elements  $A^1, A^2 \in \mathcal{B}_{2}^{\oplus}$ consist of the vector fields  $(\rho, \sigma, \tau) \longmapsto A^1 (\rho,\sigma, \tau)$ and $(\rho, \sigma, \tau) \longmapsto A^2 (\rho, \sigma, \tau)$, respectively.  The group $\g$ is unimodular and $d\mu$ is the Haar measure on it; $C_{\rho,\sigma,\tau}$ is the {\em Duflo-Moore operator} \cite{Du-Mo}, for the representation $\hat{U}^{\rho}_{\sigma,\tau}(g)$. It should also be noted that the relation (\ref{eq:ortho cond}) holds even if both the integral over  $\gd$ on the left hand side  and  the direct integral defining $\mathcal B^\oplus_2$, in the equation following, are restricted to some subset $\Sigma \subset \gd$ with non-zero Plancherel measure: $\nu_{\g}(\Sigma ) \neq 0$. In particular it holds for the set containing
 only those elements $(\rho, \sigma, \tau) \in \gd$ for which $\rho^{2}\alpha^{2}-\gamma\beta\sigma\tau\neq 0$ and none of the three parameters $\rho, \sigma, \tau$ is zero. This point will be explained further in Section \ref{subsec:wig-fcn}.

  Now restricting to the elements of $\gd$ for which  $\rho \neq 0,\; \sigma \neq 0\,\; \tau \neq 0$ and $\rho^{2}\alpha^{2}-\gamma\beta\sigma\tau\neq 0$ and denoting this set by $\mathfrak N \subset \gd$, we choose in the underlying Hilbert space $\h^\rho_{\sigma,\tau}\simeq L^{2}(\hat{\mathbb{R}}^{2},ds_{1}ds_{2})$, corresponding to the orbit labeled by $(\rho,\sigma,\tau)\in\mathfrak N$,  two measurable fields of non-zero vectors $(\rho, \sigma, \tau) \longmapsto \hat{\chi}_{\rho,\sigma,\tau}$ and $(\rho, \sigma, \tau) \longmapsto \hat{\lambda}_{\rho,\sigma,\tau}$, the field of rank-one operators $A^1(\rho, \sigma, \tau) = |\hat{\chi}_{\rho,\sigma,\tau}\rangle\langle\hat{\lambda}_{\rho,\sigma,\tau}|$ and assume that the Plancherel measure has the form $d\nu_{\g}(\rho,\sigma,\tau)=\kappa(\rho,\sigma,\tau)\;d\rho\;d\sigma\;d\tau$, for some density $\kappa$ to be determined.  Putting these into  (\ref{eq:ortho cond}) yields
\begin{eqnarray}\label{eq:ortho-con-presnt-secnario}
\langle A^{1}|A^{1}\rangle_{\mathcal{B}_{2}^{\oplus}}&=&\int_{\mathfrak N}\tr(A^{1}(\rho,\sigma,\tau)^{*}A^{1}(\rho,\sigma,\tau))\kappa(\rho,\sigma,\tau)d\rho d\sigma d\tau\nonumber\\
&=&\int_{\mathfrak N}\tr(|\hat{\lambda}_{\rho,\sigma,\tau}\rangle\langle\hat{\chi}_{\rho,\sigma,\tau}|\hat{\chi}_{\rho,\sigma,\tau}\rangle\langle\hat{\lambda}_{\rho,\sigma,\tau}|)\kappa(\rho,\sigma,\tau)d\rho d\sigma d\tau\nonumber\\
&=& \int_{\mathfrak N}\|\hat{\chi}_{\rho,\sigma,\tau}\|^{2}\|\hat{\lambda}_{\rho,\sigma,\tau}\|^{2}\kappa(\rho,\sigma,\tau)d\rho d\sigma d\tau.
\end{eqnarray}

 Since the group $\g$ is unimodular, the  Duflo-Moore operator $C_{\rho,\sigma,\tau}$ making its appearance in (\ref{eq:ortho cond}) is just a multiple of the identity operator acting on the Hilbert space $L^{2}(\mathbb{R}^{2},ds_{1}ds_{2})$.

We have the following proposition which yields the Plancherel measure and the Duflo-Moore operator of $\gd$ restricted to the sector of the unitary irreducible representations of $\g$ associated with the 4-dimensional coadjoint orbits $\mathcal{O}^{\rho,\sigma,\tau}_{4}, \; (\rho, \sigma, \tau ) \in \mathfrak N$.

\begin{prop}\label{prop:NCQM-gen-Plnchrl-DM}
The Plancherel measure of $\gd$, restricted to the sector of unitary irreducible representations $\hat{U}^{\rho}_{\sigma,\tau}, \; (\rho, \sigma, \tau) \in \mathfrak N$ (see \ref{eq:UIR-FT-gen-NCQM}) is given by
\begin{equation}\label{Plnchrl-meas-gnrl-NCQM}
d\nu_{\g}(\rho,\sigma,\tau)=\frac{|\rho^{2}\alpha^{2}-\gamma\beta\sigma\tau|}{\alpha^{2}}d\rho d\sigma d\tau
\end{equation}
and the corresponding Duflo-Moore operator reads
\begin{equation}\label{DM-gnrl-NCQM}
C_{\rho,\sigma,\tau}=(2\pi)^{\frac{5}{2}}\mathbb{I},
\end{equation}
where $\mathbb{I}$ is the identity operator defined on $L^{2}(\hat{\mathbb{R}}^{2},ds_{1}ds_{2})$.
\end{prop}

The proof is given in the Appendix.

\subsection{Construction of the Wigner function}\label{subsec:wig-fcn}
We quickly recall the construction of the Wigner function as given in \cite{plancherel,Fuhrbook}. The starting point is the  {\em operator Planceherel transform} $\mathcal P: L^2 (\g, d\mu) \longrightarrow \mathcal{B}_{2}^{\oplus}$, which for a function $f \in  L^2 (\g, d\mu )\bigcap L^1 (\g , d\mu)$ associates it to an element $A \in \mathcal{B}_{2}^{\oplus}$, given componentwise in each $\mathcal B_2 (\rho ,\sigma, \tau )$ by
\be
\mathcal P (A)(\rho, \sigma, \tau ) := A(\rho, \sigma, \tau ) = \int_{\g} f(g) \hat{U}^\rho_{\sigma , \tau}(g) C^{-1}_{\rho , \sigma , \tau}\; d\mu (g) .
\label{planch-transf}
\en
This map is an isometry in the norm
$$
  \Vert A\Vert^2_{\mathcal{B}_{2}^{\oplus}} = \int_{\g}\text{Tr} [A (\rho, \sigma, \tau )^* \; A(\rho, \sigma, \tau )]\; d\nu_{\g} \;\; , $$
i.e.,
$$ \Vert A\Vert^2_{\mathcal{B}_{2}^{\oplus}} = \Vert f\Vert^2_{L^2} = \int_{\g} \vert f(g)\vert^2\; d\mu (g), $$
which is then used to extend it to the whole of $L^2 (\g, d\mu )$.

The inverse of the map $\mathcal P$, which associates an element $A \in \mathcal{B}_{2}^{\oplus}$ to an $f \in L^2 (\g, d\mu )$
(again, first defined  on a suitable dense set and then extended by continuity) is then given by
\be
\mathcal P^{-1} (A)(g) := f(g) = \int_{\gd} \text{Tr}[ \hat{U}^\rho_{\sigma , \tau}(g)^* A(\rho, \sigma, \tau) C^{-1}_{\rho , \sigma , \tau}]\; d\nu_{\g}\; .
\label{inv-plnch-tr}
\en
It is the isometry of this map which is expressed by (\ref{eq:ortho-con-presnt-secnario}).

The next step in the construction of the Wigner function is to transform the function $f(g)$ on the group to one on its Lie algebra $\G$, using the exponential map, $g = e^X, \; X \in \G$. An $8\times 8$ matrix representation of the algebra $\G$ is given in \cite{ncqmjmp}. We write $F(X) = f(e^X)$, and also note that in the present case, $e^X = \mathbb I + X$, $\mathbb I$ being the $8 \times 8$ identity matrix. Thus, under this transformation the invariant measure $d\mu$ of $\g$ transforms to the Lebesgue measure $dX$ on $\G$. We now take the Fourier transform of $F$ to obtain a function on $\G^*$,  the dual of the Lie algebra:
$$ \widetilde{F}(X^*) = \frac 1{(2\pi )^\frac 72}\int_{\G} F(X) e^{-i\langle X^*; X\rangle}\; dX, \qquad X^* \in \G^*\; ,$$
where $\langle X^*; X\rangle$ denotes the dual pairing between $\G$ and $\G^*$. It is essentially the restriction of $\widetilde{F}(X^*)$ to a coadjoint orbit, $\mathcal O^{\rho, \sigma, \tau}$ which gives the Wigner function on that orbit.

In the present case, it is possible to write
$$ \G^* = \bigcup_{(\rho, \sigma, \tau )\in \mathbb R^3} \;\mathcal O^{\rho, \sigma, \tau}, $$
and furthermore, on the union
\be
\mathfrak g^*_0 := \bigcup_{(\rho, \sigma, \tau) \in \mathfrak N} \mathcal O^{\rho, \sigma, \tau}_4, \label{effect-orb}
\en
 of orbits comprising the set $\mathfrak N$ (recall that this is the set of all four-dimensional orbits $\mathcal O^{\rho, \sigma, \tau}_4$ with none of the parameters $\rho, \sigma, \tau$ vanishing  and $\rho^2 \alpha^2 -\gamma\beta\sigma\tau \neq 0$),  we have a splitting of the Lebesgue measure $dX^*$ of $\G^*$ in the manner
\be
   dX^* = s_{\rho, \sigma, \tau} (X^*_{\rho, \sigma, \tau }) \; d\nu_{\g}(\rho,\sigma,\tau)\; d\Omega_{\rho, \sigma, \tau}(X^*_{\rho, \sigma, \tau }), \qquad X^*_{\rho, \sigma, \tau } \in \mathcal O^{\rho, \sigma, \tau}
\label{dual-meas-split}
\en
where $s_{\rho, \sigma, \tau}$ is  a positive density, $d\Omega_{\rho, \sigma, \tau}$ is the canonical invariant measure on the coadjoint orbit $\mathcal O^{\rho, \sigma, \tau}$ (in this case, just the Lebesgue measure on $\mathbb R^4$) and  $d\nu_{\g}$ is the Plancherel measure (\ref{Plnchrl-meas-gnrl-NCQM}). Recall that the complement of the set $\mathfrak g^*_0$ has Lebesgue measure zero.

The Wigner function, defined on an orbit $\mathcal O^{\rho, \sigma, \tau}_4 \subset \mathfrak N$ and corresponding to an element $A \in \mathcal{B}_{2}^{\oplus}$  is now defined as
\bea
W(A;\; X^*_{\rho, \sigma, \tau }) &=& \frac {[s_{\rho, \sigma, \tau} (X^*_{\rho, \sigma, \tau })]^{\frac 12}}{(2\pi)^\frac 72}\widetilde{F}(X^*_{\rho, \sigma, \tau})\nonumber\\
   &=& \frac {[s_{\rho, \sigma, \tau} (X^*_{\rho, \sigma, \tau })]^{\frac 12}}{(2\pi)^\frac 72}\;\int_{\G}  e^{-i\langle X^*_{\rho, \sigma, \tau}; X\rangle}\nonumber\\
   &\times& \left[  \int_{\mathfrak N} \text{Tr}[ \hat{U}^{\omega}_{\nu , \mu}(e^{-X}) A(\omega, \nu, \mu) C^{-1}_{\omega , \nu , \mu}]\; d\nu_{\g}(\omega, \nu, \mu)\right]dX. \;\;
\label{wig-fcn-def}
\ena
Defining the direct integral Hilbert spaces
$$
   \h^\sharp = \int_{\mathfrak N}^\oplus L^2 (\mathcal O^{\rho, \sigma, \tau}_4, d\Omega_{\rho, \sigma, \tau})\; d\nu_{\g}(\rho, \sigma, \tau), \qquad \mathcal{B}^\oplus_{2}(\mathfrak N ) = \int^\oplus_{\mathfrak N} \mathcal B_2 (\rho, \sigma, \tau) \; d\nu_{\g}(\rho,\sigma,\tau),  $$
we see that the mapping $A \longmapsto W(A; \; \cdot \; ) \in \h^\sharp$ is an isometry for all $A \in \mathcal{B}^\oplus_{2}(\mathfrak N )$. Next it follows from (\ref{Plnchrl-meas-gnrl-NCQM}) that
$$ s_{\rho, \sigma, \tau} (X^*_{\rho, \sigma, \tau }) = \frac {\alpha^2}{|\rho^{2}\alpha^{2}-\gamma\beta\sigma\tau|} , $$
a constant on each orbit, and taking (\ref{DM-gnrl-NCQM}) into account, we may write
\bea
 W(A;\; X^*_{\rho, \sigma, \tau }) &=&   \frac {\vert\alpha\vert}{(2\pi)^6 \;
 \vert\rho^{2}\alpha^{2}-\gamma\beta\sigma\tau\vert^{\frac 12}} \;\int_{\G}  e^{-i\langle X^*_{\rho, \sigma, \tau}; X\rangle}\nonumber\\
   &\times& \left[  \int_{\mathfrak N} \text{Tr}[ \hat{U}^{\omega}_{\nu , \mu}(e^{-X}) A(\omega, \nu, \mu) ]\; d\nu_{\g}(\omega, \nu, \mu)\right]dX. \;\;
\label{wig-fcn-def2}
\ena
Since the group $\g$ is nilpotent, the above Wigner function is decomposable, in the sense of \cite{plancherel,Fuhrbook} (see also, Theorem \ref{Theorem:NC-Wig-func} below).  This means that if $A \in \mathcal B^\oplus_2$ is given by the operator field, $(\rho, \sigma, \tau ) \longmapsto A(\rho, \sigma, \tau ) \in \mathcal B_2 (\rho, \sigma, \tau)$, then for each $(\rho, \sigma, \tau) \in \mathfrak N$ there exists a map $W_{\rho, \sigma, \tau} : \mathcal B_2 (\rho, \sigma, \tau)  \longrightarrow L^2 (\mathcal O^{\rho, \sigma, \tau}_4, d\Omega_{\rho, \sigma, \tau})$ such that
\be
  W(A;\; X^*_{\rho, \sigma, \tau }) = [W_{\rho, \sigma, \tau} A(\rho, \sigma, \tau )] ( X^*_{\rho, \sigma, \tau }) := W (A(\rho, \sigma, \tau ); \; X^*_{\rho, \sigma, \tau }; \; \rho, \sigma, \tau),
\label{decomposition}
\en
and the norm of the function $ X^*_{\rho, \sigma, \tau } \longmapsto W (A(\rho, \sigma, \tau ); \; X^*_{\rho, \sigma, \tau }; \; \rho, \sigma, \tau)$ in the Hilbert space $L^2 (\mathcal O^{\rho, \sigma, \tau}_4, d\Omega_{\rho, \sigma, \tau})$ equals that of the norm of $A(\rho, \sigma, \tau)$ in $\mathcal B_2 (\rho, \sigma, \tau) $.
Thus  the Wigner function for the orbit $\mathcal O_4^{\rho, \sigma , \tau}$ only picks up contributions from the component of $A$, which lies in the Hilbert space $\mathcal B_2 (\rho, \sigma, \tau)$ associated to  that orbit, i.e., it is effectively the Wigner function for the operator $A(\rho, \sigma, \tau )$ in $\mathcal B_2 (\rho, \sigma, \tau) $.
The function $X^*_{\rho, \sigma, \tau} \longmapsto  W (A(\rho, \sigma , \tau); \;X^*_{\rho, \sigma, \tau}; \;\rho, \sigma , \tau )$ is thus called the Wigner function for $A(\rho, \sigma, \tau) \in \mathcal B_2 (\rho, \sigma, \tau)$, corresponding to the representation $\hat{U}^\rho_{\sigma, \tau}$ and it is defined entirely as a function on the coadjoint orbit $\mathcal O^{\rho, \sigma, \tau}_4$.

Let us now write $\rho=k_{1}$, $\sigma=k_{2}$ and $\tau=k_{3}$ satisfying $k_{1}^{2}\alpha^{2}-k_{2}k_{3}\gamma\beta\neq 0$. Denote a generic point in the orbit $\mathcal{O}^{k_{1},k_{2},k_{3}}_{4}$ by  $(k^{*}_{1},k^{*}_{2},k^{*}_{3},k^{*}_{4})$,  so that the invariant measure on this 4-dimensional coadjoint orbit reads as  $dk^{*}_{1}\;dk^{*}_{2}\;dk^{*}_{3}\;dk^{*}_{4}$. We thus have the result
\begin{thm}\label{Theorem:NC-Wig-func}
Given the Hilbert-Schmidt operator
\begin{equation*}
|\hat{\chi}_{k_1, k_2,k_3}\rangle\langle\hat{\lambda}_{k_1,k_2, k_3}|\in \mathcal B_2 (k_1, k_2, k_3) = \mathcal{B}_{2}(L^{2}(\hat{\mathbb{R}}^{2},ds_{1}ds_{2})),
\end{equation*}
with $L^{2}(\hat{\mathbb{R}}^{2},ds_{1}ds_{2})$ being the representation space for the UIR of $\g$ given by (\ref{eq:UIR-FT-gen-NCQM}), its Wigner function  (see (\ref{decomposition})) on the 4-dimensional coadjoint orbit $\mathcal{O}^{k_{1},k_{2},k_{3}}_{4}$, equipped with the Lebesgue measure $dk^{*}_{1}\;dk^{*}_{2}\;dk^{*}_{3}\;dk^{*}_{4}$, reads
\begin{align}
\MoveEqLeft W(|\hat{\chi}_{k_1,k_2,k_3}\rangle\langle\hat{\lambda}_{k_1,k_2,k_3}|;\;k_{1}^{*},k_{2}^{*},k_{3}^{*},k_{4}^{*};\;k_1,k_2,k_3)\nonumber\\
&=\frac{\vert\alpha\vert}{2\pi\vert k_{1}^{2}\alpha^{2}-k_{2}k_{3}\beta\gamma\vert^{\frac{1}{2}}}\int_{\mathbb{R}^{2}}e^{i\alpha\left(\frac{k_{1}^{*}k_{1}^{2}\alpha^{2}-k_{4}^{*}k_{1}k_{2}\alpha\beta}{k_{1}^{2}\alpha^{2}-k_{2}k_{3}\beta\gamma}\right)\widetilde{p}_{1}+i\alpha k_{2}^{*}\widetilde{p}_{2}}\nonumber\\
&\times\overline{\hat{\lambda}_{k_{1},k_{2},k_{3}}\left(\frac{1}{2}\widetilde{p}_{1}+\frac{k_{3}^{*}}{k_{1}},\frac{1}{2}\widetilde{p}_{2}+\frac{k_{1}k_{4}^{*}\alpha^{2}-k_{1}^{*}k_{3}\alpha\gamma}{k_{1}^{2}\alpha^{2}-k_{2}k_{3}\beta\gamma}\right)}\nonumber\\
&\times\hat{\chi}_{k_{1},k_{2},k_{3}}\left(-\frac{1}{2}\widetilde{p}_{1}+\frac{k_{3}^{*}}{k_{1}},-\frac{1}{2}\widetilde{p}_{2}+\frac{k_{1}k_{4}^{*}\alpha^{2}-k_{1}^{*}k_{3}\alpha\gamma}{k_{1}^{2}\alpha^{2}-k_{2}k_{3}\beta\gamma}\right)d\widetilde{p}_{1}\;d\widetilde{p}_{2}.
\label{NC-Wig-func-exct-exprssn}
\end{align}
\end{thm}

The proof is given in the Appendix.

\smallskip

\textbf{Note:}
It is to be noted that (\ref{NC-Wig-func-exct-exprssn}) can be put into a form resembling the standard Wigner function in (\ref{orig-wigfcn}) by introducing the following changes. Let us, for notational convenience, write $s_{1}$ and $s_{2}$, for $\widetilde{p}_{1}$ and $\widetilde{p}_{2}$, respectively in (\ref{NC-Wig-func-exct-exprssn}) and define the ``noncommutative positions and momenta'' by means of the 4-dimensional coadjoint orbit coordinates $(k_{1}^{*},k_{2}^{*},k_{3}^{*},k_{4}^{*}) \in \mathcal O^{k_1, k_2, k_3}_4$:
\be\label{int-vrbls-reprmtrzn}
\begin{aligned}
q_{1}^{\mathrm{nc}}&=\frac{k_{1}^{*}k_{1}^{2}\alpha^{2}-k_{4}^{*}k_{1}k_{2}\alpha\beta}{k_{1}^{2}\alpha^{2}-k_{2}k_{3}\beta\gamma},\\
q_{2}^{\mathrm{nc}}&=k_{2}^{*},\\
p_{1}^{\mathrm{nc}}&=k_{3}^{*},\\
p_{2}^{\mathrm{nc}}&=\frac{k_{1}^{2}k_{4}^{*}\alpha^{2}-k_{1}k_{1}^{*}k_{3}\alpha\gamma}{k_{1}^{2}\alpha^{2}-k_{2}k_{3}\beta\gamma}.
\end{aligned}
\en
Then (\ref{NC-Wig-func-exct-exprssn}) reads
\begin{align}
\MoveEqLeft W(A_{\rho,\sigma,\tau};\bq^{\mathrm{nc}},\bp^{\mathrm{nc}};k_1,k_2,k_3)\nonumber\\
&=\frac{\vert\alpha\vert}{2\pi\vert k_{1}^{2}\alpha^{2}-k_{2}k_{3}\beta\gamma\vert^{\frac{1}{2}}}\int_{\mathbb{R}^{2}}e^{i\alpha q_{1}^{\mathrm{nc}}s_{1}+i\alpha q_{2}^{\mathrm{nc}}s_{2}}\;\overline{\hat{\lambda}_{k_{1},k_{2},k_{3}}\left(\frac{1}{2}s_{1}+\frac{p_{1}^{\mathrm{nc}}}{k_{1}},\frac{1}{2}s_{2}+\frac{p_{2}^{\mathrm{nc}}}{k_{1}}\right)}\nonumber\\
&\times\hat{\chi}_{k_{1},k_{2},k_{3}}\left(-\frac{1}{2}s_{1}+\frac{p_{1}^{\mathrm{nc}}}{k_{1}},-\frac{1}{2}s_{2}+\frac{p_{2}^{\mathrm{nc}}}{k_{1}}\right)ds_{1}ds_{2},
\label{reprmtrzd-exprssn-wig-func}
\end{align}
which is indeed of the same form as (\ref{orig-wigfcn}), however now expressed in terms of the ''noncommutative coordinates'' $\bq^{\mathrm{nc}},\bp^{\mathrm{nc}}$. In other words, this is the Wigner function for an NCQM in which neither the two operators of position nor those of momentum commute.

\section{NC Wigner functions arising from the coadjoint orbits \texorpdfstring{$\mathcal{O}^{\rho,\sigma,0}_{4}$}{OGNC1} and \texorpdfstring{$\mathcal{O}^{\rho,0,0}_{4}$}{OGNC2} of \texorpdfstring{$\g$}{GNC}}\label{sec:wigfuncotherrep}
In this section, we attempt to throw some light on building  Wigner functions on various other 4-dimensional coadjoint orbits of $\g$, viz. $\mathcal{O}^{\rho,\sigma,0}_{4}$ and $\mathcal{O}^{\rho,0,0}_{4}$. The union of these coadjoint orbits constitute a set of Lebesgue measure zero (since they are in the complement of the set $\mathfrak g_0$  (see \ref{effect-orb})) and consequently $(\rho, \sigma, \tau) \in\!\!\!\!\!/ \;\mathfrak N$, for these orbits. We thus do not have a measure decomposition of the type  (\ref{dual-meas-split}).
Consequently, the formula  (\ref{wig-fcn-def}) to obtain the Wigner function using the Plancherel inverse formula does not  exactly fit into this case. We thus  make a slight adjustment to this definition of the Wigner function. In (\ref{wig-fcn-def2}), we consider first the exponent involving the dual pairing between the underlying Lie algebra and its dual. In the present context, the dual group parameters $\rho$, $\sigma$ and $\tau$ serve as the dual variables of the central parameters $\theta$, $\phi$ and $\psi$ of the group $\g$, respectively. Since, one has $\tau=0$ in the associated  UIR, corresponding to the 4-dimensional coadjoint orbit $\mathcal{O}^{\rho,\sigma,0}_{4}$, the central parameter $\psi$ does not make its appearance  in the explicit expression of the representation (see (2.22) of \cite{ncqmjpa}).
It therefore, points up the necessity of restricting the group manifold to a smaller region over which one integrates to obtain the Wigner function in question, because the integration with respect to the Haar measure of the full group will evidently diverge. An introduction of a factor of a delta measure $\delta(\psi)$ into the Haar measure takes care of this divergence and yields the desired Wigner function associated with the relevant family of 4-dimensional coadjoint orbits of $\g$. Following a similar argument, one needs to introduce an additional factor of $\delta(\phi)\delta(\psi)$ into the Haar measure of the group while computing the Wigner function corresponding to the other family $\mathcal{O}^{\rho,0,0}_{4}$ of the 4-dimensional coadjoint orbits of $\g$.

Let us now present our results on the NC Wigner function corresponding to the 4-dimensional coadjoint orbits $\mathcal{O}^{\rho,\sigma,0}_{4}$ by means of a theorem. The proof is omitted here as this is similar to the one used to establish theorem (\ref{Theorem:NC-Wig-func}).

\begin{thm}\label{Theorem:NC-Wig-func-1}
The Plancherel measure of $\gd$, restricted to the sector containing equivalence classes of representations of the form $\hat{U}^{\rho}_{\sigma,0}$ (see \cite{ncqmjpa}) with  $\rho \neq 0$ and $\sigma \neq 0$, is given by
\begin{equation}\label{Plnchrl-meas-gnrl-NCQM-1}
d\nu_{\g}(\rho,\sigma,\tau)=\vert\rho\vert^{2}\delta(\tau)d\rho d\sigma d\tau,
\end{equation}
and the corresponding Duflo-Moore operator reads
\begin{equation}\label{DM-gnrl-NCQM-1}
C_{\rho,\sigma}=(2\pi)^{2}\mathbb{I},
\end{equation}
where $\mathbb{I}$ is the identity operator defined on $L^{2}(\hat{\mathbb{R}}^{2},ds_{1}ds_{2})$. Also, given any Hilbert-Schmidt operator
\begin{equation*}
|\hat{\chi}_{k_1,k_2,0}\rangle\langle\hat{\lambda}_{k_1,k_2,0}|\in\mathcal{B}_{2}(k_1,k_2,0)=\mathcal{B}_{2}(L^{2}(\hat{\mathbb{R}}^{2},ds_{1}ds_{2})),
\end{equation*}
the Wigner function $W$ restricted to the 4-dimensional coadjoint orbits $\mathcal{O}^{k_{1},k_{2},0}_{4}$ equipped with the Lebesgue measure $dk^{*}_{1}\;dk^{*}_{2}\;dk^{*}_{3}\;dk^{*}_{4}$, reads
\begin{align}
\MoveEqLeft W(|\hat{\chi}_{k_1,k_2,0}\rangle\langle\hat{\lambda}_{k_1,k_2,0}|;\; k_{1}^{*},k_{2}^{*},k_{3}^{*},k_{4}^{*};\; k_1,k_2,0)\nonumber\\
&=\frac{1}{(2\pi)^{\frac{3}{2}}\vert k_{1}\vert}\int_{\mathbb{R}^{2}}e^{i\alpha\left(k_{1}^{*}-\frac{k_{4}^{*}k_{2}\beta}{k_{1}\alpha}\right)\widetilde{p}_{1}+i\alpha k_{2}^{*}\widetilde{p}_{2}}
\overline{\hat{\lambda}_{k_{1},k_{2},0}\left(\frac{1}{2}\widetilde{p}_{1}+\frac{k_{3}^{*}}{k_{1}},\frac{1}{2}\widetilde{p}_{2}+\frac{k_{4}^{*}}{k_{1}}\right)}\nonumber\\
&\times\hat{\chi}_{k_{1},k_{2},0}\left(-\frac{1}{2}\widetilde{p}_{1}+\frac{k_{3}^{*}}{k_{1}},-\frac{1}{2}\widetilde{p}_{2}+\frac{k_{4}^{*}}{k_{1}}\right)d\widetilde{p}_{1}\;d\widetilde{p}_{2}.
\label{NC-Wig-func-exct-exprssn-1}
\end{align}
\end{thm}
This then is the Wigner function for a model of NCQM in which the two momentum operators commute, but the position operators do not.

Now returning to the other family of $4$-dimensional coadjoint orbits of $\g$ that yield the standard nonrelativistic quantum mechanical representation for $2$-degrees of freedom, we sum up the results containing the respective Plancherel measure and the Wigner function in the following theorem, the proof of which, will again be omitted.

\begin{thm}\label{Theorem:NC-Wig-func-2}
The Plancherel measure of $\gd$, restricted to the sector containing equivalence classes of unitary irreducible representations of the form $\hat{U}^{\rho}_{0,0}$ (see \cite{ncqmjpa}) with $\rho \neq 0$, is given by
\begin{equation}\label{Plnchrl-meas-gnrl-NCQM-2}
d\nu_{\g}(\rho,\sigma,\tau)=\vert\rho\vert^{2}\delta(\sigma)\delta(\tau)d\rho d\sigma d\tau,
\end{equation}
while the corresponding Duflo-Moore operator reads
\begin{equation}\label{DM-gnrl-NCQM-2}
C_{\rho}=(2\pi)^{\frac{3}{2}}\mathbb{I}.
\end{equation}
Here, $\mathbb{I}$ is the identity operator defined on $L^{2}(\hat{\mathbb{R}}^{2},ds_{1}ds_{2})$. Also, given the Hilbert-Schmidt operator
\begin{equation*}
|\hat{\chi}_{k_1,0,0}\rangle\langle\hat{\lambda}_{k_1,0,0}|\in\mathcal{B}_{2}(k_1,0,0)=\mathcal{B}_{2}(L^{2}(\hat{\mathbb{R}}^{2},ds_{1}ds_{2})),
\end{equation*}
the Wigner function $W$ restricted to the 4-dimensional coadjoint orbit $\mathcal{O}^{k_{1},0,0}_{4}$ for  a fixed nonzero value of $k_{1}$ and equipped with the Lebesgue measure $dk^{*}_{1}\;dk^{*}_{2}\;dk^{*}_{3}\;dk^{*}_{4}$, reads
\begin{align}
\MoveEqLeft W(|\hat{\chi}_{k_1,0,0}\rangle\langle\hat{\lambda}_{k_1,0,0}|;k_{1}^{*},k_{2}^{*},k_{3}^{*},k_{4}^{*};k_1,0,0)\nonumber\\
&=\frac{1}{(2\pi)^{2}\vert k_{1}\vert}\int_{\mathbb{R}^{2}}e^{i\alpha(k_{1}^{*}\widetilde{p}_{1}+k_{2}^{*}\widetilde{p}_{2})}
\overline{\hat{\lambda}_{k_{1},0,0}\left(\frac{1}{2}\widetilde{p}_{1}+\frac{k_{3}^{*}}{k_{1}},\frac{1}{2}\widetilde{p}_{2}+\frac{k_{4}^{*}}{k_{1}}\right)}\nonumber\\
&\times\hat{\chi}_{k_{1},0,0}\left(-\frac{1}{2}\widetilde{p}_{1}+\frac{k_{3}^{*}}{k_{1}},-\frac{1}{2}\widetilde{p}_{2}+\frac{k_{4}^{*}}{k_{1}}\right)d\widetilde{p}_{1}\;d\widetilde{p}_{2},
\label{NC-Wig-func-exct-exprssn-2}
\end{align}
\end{thm}
which is exactly the standard Wigner function (\ref{orig-wigfcn}) of quantum mechanics.

\section{Marginals obtained from the NC Wigner function associated with the coadjoint orbits \texorpdfstring{$\mathcal{O}^{\rho,\sigma,\tau}_{4}$}{OGNC1} of \texorpdfstring{$\g$}{GNC}}\label{sec:marginals}

In this section, we  compute the marginal distributions associated to the Wigner functions (\ref{reprmtrzd-exprssn-wig-func}) arising from  the 4-dimensional coadjoint orbits $\mathcal{O}^{\rho,\sigma,\tau}_{4}$ obtained in Section (\ref{subsec:wig-fcn}). We first fix the  NC Wigner function with which we shall work using  (\ref{reprmtrzd-exprssn-wig-func}),  expressed in the noncommutative position and momentum coordinates.

\bedefin\label{def-nc-wig-func}
The NC Wigner function  arising from the group $\g$,  associated with its representation (\ref{eq:UIR-FT-gen-NCQM}) in the Landau gauge, is determined by the triple $(k_{1},k_{2},k_{3})$ satisfying $k_{1}^{2}\alpha^{2}-k_{2}k_{3}\beta\gamma\neq 0$ (see \cite{ncqmjpa} S    ection III.1 for detail) and is defined as
\be\label{Nc-wig-func-defi-eqn}
\begin{aligned}
\mathcal{W}^{\mathrm{nc}}(\bq^{\mathrm{nc}},\bp^{\mathrm{nc}};k_{1},k_{2},k_{3})&=\frac{\vert k_{1}\vert^{3}\alpha^{2}}{2\pi}W(-k_{1}\bq^{\mathrm{nc}},k_{1}\bp^{\mathrm{nc}};k_{1},k_{2},k_{3})\\
&=\frac{\vert k_{1}\alpha\vert^{3}}{(2\pi)^{2}\vert k_{1}^{2}\alpha^{2}-k_{2}k_{3}\beta\gamma\vert^{\frac{1}{2}}}\int_{\mathbb{R}^{2}}e^{-ik_{1}\alpha q_{1}^{\mathrm{nc}}s_{1}-ik_{1}\alpha q_{2}^{\mathrm{nc}}s_{2}}\\
&\times\overline{\hat{\lambda}_{k_{1},k_{2},k_{3}}\left(\frac{1}{2}s_{1}+p_{1}^{\mathrm{nc}},\frac{1}{2}s_{2}+p_{2}^{\mathrm{nc}}\right)}\\
&\times\hat{\chi}_{k_{1},k_{2},k_{3}}\left(-\frac{1}{2}s_{1}+p_{1}^{\mathrm{nc}},-\frac{1}{2}s_{2}+p_{2}^{\mathrm{nc}}\right)ds_{1}ds_{2},
\end{aligned}
\en
where $\bq^{\mathrm{nc}}$ and $\bp^{\mathrm{nc}}$ are as given by (\ref{int-vrbls-reprmtrzn}).
\findefi
\begin{remark}\label{remrk-wig-func}
We remark here  that as $k_{2},k_{3}\rightarrow 0$, (\ref{Nc-wig-func-defi-eqn}) reproduces the canonical Wigner function for the 2-dimensional Weyl-Heisenberg group associated with standard quantum mechanics (see (\ref{orig-wigfcn})):
\bea\label{lim-stndard-qm}
\lefteqn{\lim_{\substack{k_{2}\to 0 \\ k_{3}\to 0}}\mathcal{W}^{\mathrm{nc}}(\bq^{\mathrm{nc}},\bp^{\mathrm{nc}};k_{1},k_{2},k_{3})}\nn\\
&&=W^{\mathrm{QM}}(k_{1}^{*},k_{2}^{*},k_{3}^{*},k_{4}^{*};\hbar)\nn\\
&&=\frac{1}{4\pi^{2}\hbar^{2}}\int_{\mathbb{R}^{2}}e^{-\frac{i}{\hbar}k_{1}^{*}s_{1}-\frac{i}{\hbar}k_{2}^{*}s_{2}}
\overline{\hat{\lambda}_{\hbar}\left(\frac{1}{2}s_{1}+k_{3}^{*},\frac{1}{2}s_{2}+k_{4}^{*}\right)}\nn\\
&&\times\hat{\chi}_{\hbar}\left(-\frac{1}{2}s_{1}+k_{3}^{*},-\frac{1}{2}s_{2}+k_{4}^{*}\right)ds_{1}ds_{2},
\ena
where we chose $\hbar=\frac{1}{k_{1}\alpha}$ and denoted the $L^{2}$-functions $\hat{\lambda}_{k_{1},0,0}$ and $\hat{\chi}_{k_{1},0,0}$ simply by $\hat{\lambda}_{\hbar}$ and $\hat{\chi}_{\hbar}$, respectively.
\end{remark}

If we integrate out $\bq^{\mathrm{nc}}$ or $\bp^{\mathrm{nc}}$ from the quasi-probability distribution  $\displaystyle\mathcal{W}^{\mathrm{nc}}(\bq^{\mathrm{nc}},\bp^{\mathrm{nc}}\allowbreak;k_{1},k_{2},k_{3})$, we obtain the classical probability density function (upto a non-vanishing constant factor) called the marginal distribution for the noncommutative momentum or the position, respectively. Also, for notational convenience, since the $L^{2}$-functions, say, $\hat{\chi}_{k_{1},k_{2},k_{3}}$, we have been considering, are all supported on the respective coadjoint orbits determined by $(k_1,k_2,k_3)$, we will be dropping the subscripts without loss of any generality.

The noncommutative marginal probability distribution in momentum coordinates $\bp^{\mathrm{nc}}$ due to a rank one operator $\displaystyle\vert\hat{\psi}\rangle\langle\hat{\psi}\vert$ is given by
\be\label{eq-thm-margnl-momntm}
\int_{\mathbb{R}^{2}}\mathcal{W}^{\mathrm{nc}}(\bq^{\mathrm{nc}},\bp^{\mathrm{nc}};k_{1},k_{2},k_{3})d\bq^{\mathrm{nc}}=\frac{\vert k_{1}\alpha\vert}{\vert k_{1}^{2}\alpha^{2}-k_{2}k_{3}\beta\gamma\vert^{\frac 12}}\vert\hat{\psi}(\bp^{\mathrm{nc}})\vert^{2},
\en

Expressing $\hat{\psi}\in L^{2}(\mathbb{R}^{2},d\bs)$ by its inverse Fourier transform $\psi\in L^{2}(\mathbb{R}^{2},d\br)$ in (\ref{Nc-wig-func-defi-eqn}), the NC Wigner function for the rank one operator $\vert\psi\rangle\langle\psi\vert$ reads
\bea\label{rewrt-wig-func-nc}
\lefteqn{\mathcal{W}^{\mathrm{nc}}(\bq^{\mathrm{nc}},\bp^{\mathrm{nc}};k_{1},k_{2},k_{3})}\nonumber\\
&&=\frac{\vert k_{1}\alpha\vert^{3}}{(2\pi)^{2}\vert k_{1}^{2}\alpha^{2}-k_{2}k_{3}\beta\gamma\vert^{\frac{1}{2}}}e^{ik_{1}\alpha \bp^{\mathrm{nc}}.\br}\overline{\psi\left(\frac{1}{2}\br+\bq^{\mathrm{nc}}\right)}\psi\left(-\frac{1}{2}\br+\bq^{\mathrm{nc}}\right)d\br.
\ena
Hence the noncommutative marginal probability distribution in the position coordinates $\bq^{\mathrm{nc}}$ for the rank one operator $\displaystyle\vert\psi\rangle\langle\psi\vert$ can easily be read off as
\be\label{eq-thm-margnl-postn}
\int_{\mathbb{R}^{2}}\mathcal{W}^{\mathrm{nc}}(\bq^{\mathrm{nc}},\bp^{\mathrm{nc}};k_{1},k_{2},k_{3})d\bp^{\mathrm{nc}}=\frac{\vert k_{1}\alpha\vert}{\vert k_{1}^{2}\alpha^{2}-k_{2}k_{3}\beta\gamma\vert^{\frac 12}}\vert\psi(\bq^{\mathrm{nc}})\vert^{2}.
\en

Let us express the dimensionful noncommutativity parameters $\hbar$, $\vartheta$ and $\mathcal{B}$ using the triple $(k_1,k_2,k_3)$ that fixes the coadjoint orbit under study:
\be\label{nc-parametrs}
\begin{aligned}
\hbar&=\frac{1}{k_{1}\alpha},\\
\vartheta&=-\frac{k_{2}\beta}{k_{1}^{2}\alpha^{2}},\\
\mathcal{B}&=-\frac{k_{3}\gamma}{k_{1}^{2}\alpha^{2}}.
\end{aligned}
\en
We have already introduced the canonical coordinates of the 4-dimensional coadjoint orbits by $(k_{1}^{*},k_{2}^{*},k_{3}^{*},k_{4}^{*})$. Using these coordinates, the noncommutativity parameters introduced in (\ref{nc-parametrs}) and the equations (\ref{rewrt-wig-func-nc}) and (\ref{int-vrbls-reprmtrzn}), the NC Wigner function (see definition (\ref{def-nc-wig-func})) for the rank one operator $\vert\psi\rangle\langle\psi\vert$ reads
\bea\label{nc-wig-func-nc-prmtrs}
\lefteqn{\mathcal{W}^{\mathrm{nc}}(k_{1}^{*},k_{2}^{*},k_{3}^{*},k_{4}^{*};\hbar,\vartheta,\mathcal{B})}\nn\\
&&=\frac{1}{4\pi^{2}\vert\hbar\vert\sqrt{\vert\hbar^{2}-\mathcal{B}\vartheta\vert}}\int_{\mathbb{R}^{2}}e^{\frac{i}{\hbar}k_{3}^{*}r_{1}+\frac{i(\hbar k_{4}^{*}+\mathcal{B}k_{1}^{*})}{\hbar^{2}-\mathcal{B}\vartheta}r_{2}}\nn\\
&&\times\overline{\psi\left(\frac{1}{2}r_{1}+\frac{\hbar^{2} k_{1}^{*}+\hbar\vartheta k_{4}^{*}}{\hbar^{2}-\mathcal{B}\vartheta},\frac{1}{2}r_{2}+k_{2}^{*}\right)}\nn\\
&&\times\psi\left(-\frac{1}{2}r_{1}+\frac{\hbar^{2} k_{1}^{*}+\hbar\vartheta k_{4}^{*}}{\hbar^{2}-\mathcal{B}\vartheta},-\frac{1}{2}r_{2}+k_{2}^{*}\right)dr_{1}dr_{2},
\ena
where $\psi\in L^{2}(\mathbb{R}^{2},d\br)$.

Now let us recall that the marginal distributions associated with the NC Wigner functions were computed by integrating out the noncommutative momenta or the positions in (\ref{eq-thm-margnl-momntm},\ref{eq-thm-margnl-postn}). These marginals were found to be probability distributions (upto a positive scaling factor) and hence, strictly positive, in the respective noncommutative positions or momenta (see \ref{int-vrbls-reprmtrzn}). In the current literature (see \cite{bastoscmp} and \cite{wigfuncchinese}, for example), the noncommutative marginal distributions are defined to be obtained by integrating the NC Wigner functions with respect to the canonical coordinates, i.e. the coadjoint orbit coordinates  $k_{1}^{*}$, $k_{2}^{*}$, $k_{3}^{*}$ and $k_{4}^{*}$. The marginals, thus obtained are not strictly positive and can be expressed by means of certain $\star$-products (see \cite{bastoscmp}, for example). Let us quickly verify if the NC Wigner function (\ref{nc-wig-func-nc-prmtrs}) arising from the group $\g$ produces similar marginals as in (\cite{bastoscmp},\cite{wigfuncchinese}) when integrated with respect to the canonical coordinates of the underlying 4-dimensional coadjoint orbit.

Given two Wigner functions $W(\vert\chi\rangle\langle\lambda\vert; k_{1}^{\star},k_{2}^{\star},k_{3}^{\star},k_{4}^{\star})$ and $W(\vert\phi\rangle\langle\psi\vert; k_{1}^{\star},k_{2}^{\star},k_{3}^{\star},k_{4}^{\star})$ associated with the rank one operators $\vert\chi\rangle\langle\lambda\vert$ and $\vert\phi\rangle\langle\psi\vert$, respectively, their multiplication does not necessarily yield a square integrable function on the underlying coadjoint orbit. It is, therefore, necessary to deform the usual multiplication of the Wigner functions associated to a fixed coadjoint orbit to obtain a square integrable function on it. In the following we define the various $\star$-products for the Wigner functions obtained in (\ref{nc-wig-func-nc-prmtrs}).

\bedefin\label{def-star-prod-nc-wig-func}
Let us first recall that a coadjoint orbit, for the group $\g$  is determined by a triple $(\hbar,\vartheta,\mathcal{B})$. When none of $\frac{1}{\hbar}$, $\vartheta$ and $\mathcal{B}$ is zero with $\hbar^{2}-\mathcal{B}\vartheta\neq 0$, the underlying orbit is 4-dimensional; its coordinates being given by $(k_{1}^{*},k_{2}^{*},k_{3}^{*},k_{4}^{*})$. Here, $k_{1}^{*}$ and $k_{2}^{*}$ stand for the position coordinates, while $k_{3}^{*}$ and $k_{4}^{*}$ for the respective momenta.

If we choose the momentum coordinates $k_{3}^{*}=p_1$ and $k_{4}^{*}=p_2$ to be fixed apriori, then the kernel representation of the $\star_{\vartheta}$-product is defined as
\bea\label{def:star-vartheta}
\lefteqn{\mathcal{W}^{\mathrm{nc}}(\vert\chi\rangle\langle\lambda\vert;k_{1}^{*},k_{2}^{*},p_1,p_2)\star_{\vartheta}\mathcal{W}^{\mathrm{nc}}(\vert\phi\rangle\langle\psi\vert;k_{1}^{*},k_{2}^{*},p_{1},p_{2})}\nn\\
&&=\frac{\sqrt{\vert\hbar^{2}-\mathcal{B}\vartheta\vert}}{\pi \vert\hbar\vartheta\vert}\int_{\mathbb{R}^{2}}\exp{\frac{i}{\vartheta}\left[\begin{smallmatrix}k_{1}^{*}-\eta_{1}&k_{2}^{*}-\eta_{2}\end{smallmatrix}\right]\left[\begin{smallmatrix}0&1\\-1&0\end{smallmatrix}\right]\left[\begin{smallmatrix}\eta_{1}-k_{1}^{*}\\k_{2}^{*}-\eta_{2}\end{smallmatrix}\right]}\nn\\
&&\times\mathcal{W}^{\mathrm{nc}}(\vert\chi\rangle\langle\lambda\vert;\eta_1,\eta_2,p_1,p_2)\mathcal{W}^{\mathrm{nc}}(\vert\phi\rangle\langle\psi\vert;\eta_1,2k_{2}^{*}-\eta_2,p_1,p_2)d\eta_{1}d\eta_{2}.\nn\\
\ena

If one chooses the position coordinates $k_{1}^{*}=q_1$ and $k_{2}^{*}=q_2$ to be fixed instead, then the kernel representation of the $\star_{\mathcal{B}}$-product is defined as
\bea\label{def:star-mathcalB}
\lefteqn{\mathcal{W}^{\mathrm{nc}}(\vert\chi\rangle\langle\lambda\vert;q_1,q_{2},k_{3}^{*},k_{4}^{*})\star_{\mathcal{B}}\mathcal{W}^{\mathrm{nc}}(\vert\phi\rangle\langle\psi\vert;q_1,q_2,k_{3}^{*},k_{4}^{*})}\nn\\
&&=\frac{\sqrt{\vert\hbar^{2}-\mathcal{B}\vartheta\vert}}{\pi \vert\hbar\mathcal{B}\vert}\int_{\mathbb{R}^{2}}\exp{\frac{i}{\mathcal{B}}\left[\begin{smallmatrix}k_{3}^{*}-\xi_{1}&k_{4}^{*}-\xi_{2}\end{smallmatrix}\right]\left[\begin{smallmatrix}0&1\\-1&0\end{smallmatrix}\right]\left[\begin{smallmatrix}k_{3}^{*}-\xi_{1}\\\xi_{2}-k_{4}^{*}\end{smallmatrix}\right]}\nn\\
&&\times\mathcal{W}^{\mathrm{nc}}(\vert\chi\rangle\langle\lambda\vert;q_1,q_2,\xi_1,\xi_2)\mathcal{W}^{\mathrm{nc}}(\vert\phi\rangle\langle\psi\vert;q_1,q_2,2k_{3}^{*}-\xi_1,\xi_2)d\xi_{1}d\xi_{2}.\nn\\
\ena

Between two Wigner functions corresponding to the coadjoint orbit with coordinates given by $(k_{1}^{*},k_{2}^{*},k_{3}^{*},k_{4}^{*})$, one can define the kernel representation of the $\star_{\hbar}$-product in the following way

\bea\label{def:star-hbar}
\lefteqn{\mathcal{W}^{\mathrm{nc}}(\vert\chi\rangle\langle\lambda\vert;k_{1}^{*},k_{2}^{*},k_{3}^{*},k_{4}^{*})\star_{\hbar}\mathcal{W}^{\mathrm{nc}}(\vert\phi\rangle\langle\psi\vert;k_{1}^{*},k_{2}^{*},k_{3}^{*},k_{4}^{*})}\nn\\
&&=\frac{\sqrt{\vert\hbar^{2}-\mathcal{B}\vartheta\vert}}{\pi \vert\hbar\vert}\int_{\mathbb{R}^{2}}\int_{\mathbb{R}^{2}}\exp{\frac{i}{\hbar}\left[\begin{smallmatrix}k_{1}^{*}-\eta_{1}&k_{2}^{*}-\eta_{2}&k_{3}^{*}-\xi_{1}&k_{4}^{*}-\xi_{2}\end{smallmatrix}\right]\left[\begin{smallmatrix}0&0&1&0\\0&0&0&1\\-1&0&0&0\\0&-1&0&0\end{smallmatrix}\right]\left[\begin{smallmatrix}\eta_{1}-k_{1}^{*}\\k_{2}^{*}-\eta_{2}\\k_{3}^{*}-\xi_{1}\\\xi_{2}-k_{4}^{*}\end{smallmatrix}\right]}\nn\\
&&\times\mathcal{W}^{\mathrm{nc}}(\vert\chi\rangle\langle\lambda\vert;\eta_1,\eta_2,\xi_1,\xi_2)\mathcal{W}^{\mathrm{nc}}(\vert\phi\rangle\langle\psi\vert;\eta_1,2k_{2}^{*}-\eta_{2},2k_{3}^{*}-\xi_{1},\xi_2)d\eta_{1}d\eta_{2}d\xi_{1}d\xi_{2}.\nn\\
\ena

Combining (\ref{def:star-hbar}), (\ref{def:star-vartheta}) and (\ref{def:star-mathcalB}), one can define a more general $\star$-product denoted with $\star_{\scriptscriptstyle{\hbar,\vartheta,\mathcal{B}}}$, the kernel representation of which is given by

\bea\label{def:star-prod-gen}
\lefteqn{\mathcal{W}^{\mathrm{nc}}(\vert\chi\rangle\langle\lambda\vert;k_{1}^{*},k_{2}^{*},k_{3}^{*},k_{4}^{*})\star_{\scriptscriptstyle{\hbar,\vartheta,\mathcal{B}}}\mathcal{W}^{\mathrm{nc}}(\vert\phi\rangle\langle\psi\vert;k_{1}^{*},k_{2}^{*},k_{3}^{*},k_{4}^{*})}\nn\\
&&=\frac{\sqrt{\vert\hbar^{2}-\mathcal{B}\vartheta\vert}}{\pi \vert\hbar\vert}\int_{\mathbb{R}^{2}}\int_{\mathbb{R}^{2}}\exp{\frac{i}{\hbar^{2}-\mathcal{B}\vartheta}\left[\begin{smallmatrix}k_{1}^{*}-\eta_{1}&k_{2}^{*}-\eta_{2}&k_{3}^{*}-\xi_{1}&k_{4}^{*}-\xi_{2}\end{smallmatrix}\right]\left[\begin{smallmatrix}0&\mathcal{B}&-\hbar&0\\-\mathcal{B}&0&0&-\hbar\\\hbar&0&0&\vartheta\\0&\hbar&-\vartheta&0\end{smallmatrix}\right]\left[\begin{smallmatrix}\eta_{1}-k_{1}^{*}\\k_{2}^{*}-\eta_{2}\\k_{3}^{*}-\xi_{1}\\\xi_{2}-k_{4}^{*}\end{smallmatrix}\right]}\nn\\
&&\times\mathcal{W}^{\mathrm{nc}}(\vert\chi\rangle\langle\lambda\vert;\eta_1,\eta_2,\xi_1,\xi_2)\mathcal{W}^{\mathrm{nc}}(\vert\phi\rangle\langle\psi\vert;\eta_{1},2k_{2}^{*}-\eta_{2},2k_{3}^{*}-\xi_{1},\xi_{2})d\eta_{1}d\eta_{2}d\xi_{1}d\xi_{2}.\nn\\
\ena
\findefi

Inspired by the definitions provided in (\ref{def-star-prod-nc-wig-func}, one can compute the noncommutative marginal distributions in terms of $\star_{\vartheta}$ and $\star_{\mathcal{B}}$ products:

\begin{proposition}\label{prop:marginal-dstrbn}
The NC Wigner function (\ref{nc-wig-func-nc-prmtrs})of a rank one operator $\vert\psi\rangle\langle\psi\vert$ can be integrated with respect to the momentum coordinates $k_{3}^{*}$ and $k_{4}^{*}$ to yield the following noncommutative marginal distribution in position coordinates:
\be\label{eq:nc-marg-pos-star}
\int_{\mathbb{R}^{2}}\mathcal{W}^{\mathrm{nc}}(\vert\psi\rangle\langle\psi\vert ;k_{1}^{*},k_{2}^{*},k_{3}^{*},k_{4}^{*})dk_{3}^{*}dk_{4}^{*}=\overline{\psi(k_{1}^{*},k_{2}^{*})}\star_{\vartheta}\psi(k_{1}^{*},k_{2}^{*}),
\en
where $\psi\in L^{2}(\mathbb{R}^{2},d\br)$.

Also, the noncommutative marginal distribution in momentum coordinates is given by
\be\label{eq:nc-marg-momntm-star}
\int_{\mathbb{R}^{2}}\mathcal{W}^{\mathrm{nc}}(\vert\hat{\psi}\rangle\langle\hat{\psi}\vert ;k_{1}^{*},k_{2}^{*},k_{3}^{*},k_{4}^{*})dk_{1}^{*}dk_{2}^{*}=\overline{\hat{\psi}(k_{3}^{*},k_{4}^{*})}\star_{\mathcal{B}}\hat{\psi}(k_{3}^{*},k_{4}^{*}),
\en
where $\hat{\psi}\in L^{2}(\mathbb{R}^{2},d\bs)$.
\end{proposition}

\prf{.}
Keeping the position coordinates $k_{1}^{*}$, $k_{2}^{*}$ fixed
in (\ref{nc-wig-func-nc-prmtrs}), one can integrate out the momenta coordinates to yield

\bea\label{verfy-litrte}
\lefteqn{\int_{\mathbb{R}^{2}}\mathcal{W}^{\mathrm{nc}}(k_{1}^{*},k_{2}^{*},k_{3}^{*},k_{4}^{*};\hbar,\vartheta,\mathcal{B})dk_{3}^{*}dk_{4}^{*}}\nn\\
&&=\frac{1}{4\pi^{2}\vert\hbar\vert\sqrt{\vert \hbar^{2}-\mathcal{B}\vartheta\vert}}\int_{\mathbb{R}^{2}}\int_{\mathbb{R}^{2}}e^{\frac{i}{\hbar}k_{3}^{*}r_{1}+\frac{i(\hbar k_{4}^{*}+\mathcal{B}k_{1}^{*})}{\hbar^{2}-\mathcal{B}\vartheta}r_{2}}\nn\\
&&\times\overline{\psi\left(\frac{1}{2}r_{1}+\frac{\hbar^{2} k_{1}^{*}+\hbar\vartheta k_{4}^{*}}{\hbar^{2}-\mathcal{B}\vartheta},\frac{1}{2}r_{2}+k_{2}^{*}\right)}\nn\\
&&\times\psi\left(-\frac{1}{2}r_{1}+\frac{\hbar^{2} k_{1}^{*}+\hbar\vartheta k_{4}^{*}}{\hbar^{2}-\mathcal{B}\vartheta},-\frac{1}{2}r_{2}+k_{2}^{*}\right)dr_{1}dr_{2}dk_{3}^{*}dk_{4}^{*}\nn\\
&&=\frac{1}{2\pi\sqrt{\vert\hbar^{2}-\mathcal{B}\vartheta\vert}}\int_{\mathbb{R}^{2}}e^{\frac{i(\hbar k_{4}^{*}+\mathcal{B}k_{1}^{*})}{\hbar^{2}-\mathcal{B}\vartheta}r_{2}}\overline{\psi\left(\frac{\hbar^{2} k_{1}^{*}+\hbar\vartheta k_{4}^{*}}{\hbar^{2}-\mathcal{B}\vartheta},\frac{1}{2}r_{2}+k_{2}^{*}\right)}\nn\\
&&\times\psi\left(\frac{\hbar^{2} k_{1}^{*}+\hbar\vartheta k_{4}^{*}}{\hbar^{2}-\mathcal{B}\vartheta},-\frac{1}{2}r_{2}+k_{2}^{*}\right)dr_{2}dk_{4}^{*}\nn\\
&&=\frac{\sqrt{\vert\hbar^{2}-\mathcal{B}\vartheta\vert}}{\pi \vert\hbar\vartheta\vert}\int_{\mathbb{R}^{2}}e^{\frac{2i}{\vartheta}(\eta_{1}-k_{1}^{*})(\eta_{2}-k_{2}^{*})}\overline{\psi(\eta_{1},\eta_{2})}\psi(\eta_{1},2k_{2}^{*}-\eta_{2})d\eta_{1}\;d\eta_{2}\nn\\
&&=\frac{\sqrt{\vert\hbar^{2}-\mathcal{B}\vartheta\vert}}{\pi\vert \hbar\vartheta\vert}\int_{\mathbb{R}^{2}}\exp{\frac{i}{\vartheta}\begin{bmatrix}k_{1}^{*}-\eta_{1}&k_{2}^{*}-\eta_{2}\end{bmatrix}\begin{bmatrix}0&1\\-1&0\end{bmatrix}\begin{bmatrix}\eta_{1}-k_{1}^{*}\\k_{2}^{*}-\eta_{2}\end{bmatrix}}\nn\\
&&\times\overline{\psi(\eta_{1},\eta_{2})}\psi(\eta_{1},2k_{2}^{*}-\eta_{2})d\eta_{1}d\eta_{2}\nn\\
&&=\overline{\psi(k_{1}^{*},k_{2}^{*})}\star_{\vartheta}\psi(k_{1}^{*},k_{2}^{*}),
\ena
where in the last line we have exploited the definition of the kernel representation of the $\star_{\vartheta}$-product given by (\ref{def:star-vartheta}). Also, proceeding to third line from the second above is justified by means of the following change of variables:
\be\label{eqn-chng-of-vrbl-star-prod}
\begin{aligned}
\frac{\hbar^{2}k_{1}^{*}+\hbar\vartheta k_{4}^{*}}{\hbar^{2}-\mathcal{B}\vartheta}&=\eta_{1},\\
\frac{1}{2}r_{2}+k_{2}^{*}&=\eta_{2}.
\end{aligned}
\en

Now considering the momentum space representation of the NC Wigner function, one obtains for $\vert\hat{\psi}\rangle\langle\hat{\psi}\vert$:
\bea\label{verfy-litrte-momntm-rep}
\lefteqn{\int_{\mathbb{R}^{2}}\mathcal{W}^{\mathrm{nc}}(k_{1}^{*},k_{2}^{*},k_{3}^{*},k_{4}^{*};\hbar,\vartheta,\mathcal{B})dk_{1}^{*}dk_{2}^{*}}\nn\\
&&=\frac{1}{4\pi^{2}\vert\hbar\vert\sqrt{\vert\hbar^{2}-\mathcal{B}\vartheta\vert}}\int_{\mathbb{R}^{2}}\int_{\mathbb{R}^{2}}e^{-i\left(\frac{\hbar k_{1}^{*}+\vartheta k_{4}^{*}}{\hbar^{2}-\mathcal{B}\vartheta}\right)s_{1}-\frac{i}{\hbar}k_{2}^{*}s_{2}}\nn\\
&&\times\overline{\hat{\psi}\left(\frac{1}{2}s_{1}+k_{3}^{*},\frac{1}{2}s_{2}+\frac{\hbar^{2} k_{4}^{*}+\hbar\mathcal{B} k_{1}^{*}}{\hbar^{2}-\mathcal{B}\vartheta}\right)}\nn\\
&&\times\hat{\psi}\left(-\frac{1}{2}s_{1}+k_{3}^{*},-\frac{1}{2}s_{2}+\frac{\hbar^{2} k_{4}^{*}+\hbar\mathcal{B} k_{1}^{*}}{\hbar^{2}-\mathcal{B}\vartheta}\right)ds_{1}ds_{2}dk_{1}^{*}dk_{2}^{*}\nn\\
&&=\frac{1}{2\pi\sqrt{\vert\hbar^{2}-\mathcal{B}\vartheta\vert}}\int_{\mathbb{R}^{2}}e^{-i\left(\frac{\hbar k_{1}^{*}+\vartheta k_{4}^{*}}{\hbar^{2}-\mathcal{B}\vartheta}\right)s_{1}}\overline{\hat{\psi}\left(\frac{1}{2}s_{1}+k_{3}^{*},\frac{\hbar^{2} k_{4}^{*}+\hbar\mathcal{B} k_{1}^{*}}{\hbar^{2}-\mathcal{B}\vartheta}\right)}\nn\\
&&\times\hat{\psi}\left(-\frac{1}{2}s_{1}+k_{3}^{*},\frac{\hbar^{2} k_{4}^{*}+\hbar\mathcal{B} k_{1}^{*}}{\hbar^{2}-\mathcal{B}\vartheta}\right)ds_{1}dk_{1}^{*}\nn\\
&&=\frac{\sqrt{\vert\hbar^{2}-\mathcal{B}\vartheta\vert}}{\pi\vert \hbar\mathcal{B}\vert}\int_{\mathbb{R}^{2}}e^{-\frac{2i}{\mathcal{B}}(\xi_{2}-k_{4}^{*})(\xi_{1}-k_{3}^{*})}\overline{\hat{\psi}(\xi_1,\xi_2)}\hat{\psi}(2k_{3}^{*}-\xi_{1},\xi_{2})d\xi_{1}d\xi_{2}\nn\\
&&=\frac{\sqrt{\vert\hbar^{2}-\mathcal{B}\vartheta\vert}}{\pi \vert\hbar\mathcal{B}\vert}\int_{\mathbb{R}^{2}}\exp{\frac{i}{\mathcal{B}}\begin{bmatrix}k_{3}^{*}-\xi_{1}&k_{4}^{*}-\xi_{2}\end{bmatrix}\begin{bmatrix}0&1\\-1&0\end{bmatrix}\begin{bmatrix}k_{3}^{*}-\xi_{1}\\\xi_{2}-k_{4}^{*}\end{bmatrix}}\nn\\
&&\times\overline{\hat{\psi}(\xi_{1},\xi_{2})}\hat{\psi}(2k_{3}^{*}-\xi_{1},\xi_{2})d\xi_{1}d\xi_{2}\nn\\
&&=\overline{\hat{\psi}(k_{3}^{*},k_{4}^{*})}\star_{\mathcal{B}}\hat{\psi}(k_{3}^{*},k_{4}^{*}).
\ena
In the last line above, we have used the definition (\ref{def:star-mathcalB}) of the kernel representation of the $\star_{\mathcal{B}}$-product. Also, going to the third line from the second was justified by the following change of variables:
\be\label{eqn-chng-of-vrbl-star-prod-mom-rep}
\begin{aligned}
\frac{1}{2}s_{1}+k_{3}^{*}&=\xi_{1},\\
\frac{\hbar^{2}k_{4}^{*}+\hbar\mathcal{B} k_{1}^{*}}{\hbar^{2}-\mathcal{B}\vartheta}&=\xi_{2}.
\end{aligned}
\en
\qed

\section{Conclusion and future perspectives}\label{sec:conclusion}
We re-emphasize here that our approach to the construction of the different Wigner functions is based on the representations of the group $\g$, which we consider to be the kinematical symmetry group of noncommutative quantum mechanics. This is in contrast to the approach adopted in \cite{bastoscmp}, for example. Since by using our method, we have been able to obtain Wigner functions for theories with different levels of non-commutativity, using representations of the same group, it tends to demonstrate the versatility of the approach. Indeed, the representations of $\g$ which we have used to construct our Wigner functions, include models of NCQM in which, apart from the canonical commutation relations between position and momentum, (1) neither the two operators of position, nor those of momentum commute, (2)  the two momenta commute but the positions do not and (3) both the two momenta and the two position operators commute (standard quantum mechanics). It is indeed remarkable that all these representations come from the same group.

In (\cite{nctori}), a 2-parameter family of equivalent unitary irreducible representations (for fixed $\rho$, $\sigma$ and $\tau$) of the group $\g$ has been constructed. For certain values of the two parameters involved there, one precisely obtains the representations corresponding to the Landau and the symmetric gauges of NCQM. In this paper, we have exploited the Landau gauge representation of $\g$ (see \ref{eq:UIR-gen-NCQM}) to construct its various Wigner functions and defined the kernel representations of the associated $\star$-products between them. It would be interesting to see how the Wigner functions and the associated $\star$-products, defined on the respective coadjoint orbits, turn out to be for  other members of the 2-parameter family of gauge equivalent representations, e.g., the symmetric gauge representation of $\g$ (see p.19, \cite{ncqmjpa}).

\section{Appendix}\label{sec-app}
In this Appendix we collect together the proofs of some of the results quoted in the paper.

\prf {\bf of Proposition \ref{prop:NCQM-gen-Plnchrl-DM}}

We start out by observing that subject to $A^{1}(\rho,\sigma,\tau)=A^{2}(\rho,\sigma,\tau)=|\hat{\chi}_{\rho,\sigma,\tau}\rangle\langle\hat{\lambda}_{\rho,\sigma,\tau}|$ in (\ref{eq:ortho cond}) and the fact that the underlying Duflo-Moore operator $C_{\rho,\sigma,\tau}=N\mathbb{I}$ with $N$ being a real number and $\mathbb{I}$ being the identity operator acting on the Hilbert space $L^{2}(\hat{\mathbb{R}}^{2},ds_{1}ds_{2})$, the left side of (\ref{eq:ortho cond}) now reads
\begin{eqnarray}\label{eq:planchrl-dervtn}
\lefteqn{\frac{1}{N^2}\int_{\mathbb{R}^7}\left[\int_{\mathbb{R}^{*}\times\mathbb{R}^{*}\times\mathbb{R}^{*}}\overline{\langle\hat{\chi}_{\rho,\sigma,\tau}|\hat{U}^{\rho}_{\sigma,\tau}(\theta,\phi,\psi,\bq,\bp)\hat{\lambda}_{\rho,\sigma,\tau}\rangle}\kappa(\rho,\sigma,\tau)d\rho d\sigma d\tau\right.}\nonumber\\
&&\left.\times\int_{\mathbb{R}^{*}\times\mathbb{R}^{*}\times\mathbb{R}^{*}}\langle\hat{\chi}_{\rho^{\prime},\sigma^{\prime},\tau^{\prime}}|\hat{U}^{\rho^{\prime}}_{\sigma^{\prime},\tau^{\prime}}(\theta,\phi,\psi,\bq,\bp)\hat{\lambda}_{\rho^{\prime},\sigma^{\prime},\tau^{\prime}}\rangle\kappa(\rho^{\prime},\sigma^{\prime},\tau^{\prime})d\rho^{\prime}d\sigma^{\prime}d\tau^{\prime}\right]\nonumber\\
&&\times d\theta\; d\phi\; d\psi\; d\bq\; d\bp\nonumber\\
=\lefteqn{\frac{1}{N^2}\int_{(\rho,\sigma,\tau)}\int_{(\rho^{\prime},\sigma^{\prime},\tau^{\prime})}\left[\int_{\mathbb{R}^{7}}\left\{\int_{(s_{1},s_{2})\in\mathbb{R}^{2}}\int_{(s_{1}^{\prime},s_{2}^{\prime})\in\mathbb{R}^{2}}e^{i(\rho-\rho^{\prime})\theta+i(\sigma-\sigma^{\prime})\phi+i(\tau-\tau^{\prime})\psi}\right.\right.}\nonumber\\
&&\left.\left.\times e^{i\alpha q_{1}(\rho s_{1}-\rho^{\prime}s_{1}^{\prime})+i\alpha q_{2}(\rho s_{2}-\rho^{\prime}s_{2}^{\prime})+i\beta p_{1}(\sigma s_{2}-\sigma^{\prime}s_{2}^{\prime})-\frac{i\alpha}{2}(\rho-\rho^{\prime})q_{1}p_{1}-\frac{i\alpha}{2}(\rho-\rho^{\prime})q_{2}p_{2}}\right.\right.\nonumber\\
&&\left.\left.\times e^{-\frac{i\gamma}{2}(\tau-\tau^{\prime})q_{1}q_{2}-\frac{i\beta}{2}(\sigma-\sigma^{\prime})p_{1}p_{2}}\overline{\hat{\lambda}_{\rho,\sigma,\tau}\left(s_{1}-\frac{\tau\gamma}{\rho\alpha}q_{2}-p_{1},s_{2}-p_{2}\right)}\hat{\chi}_{\rho,\sigma,\tau}(s_{1},s_{2})\right.\right.\nonumber\\
&&\left.\left.\times\overline{\hat{\chi}_{\rho^{\prime},\sigma^{\prime},\tau^{\prime}}(s^{\prime}_{1},s^{\prime}_{2})}\hat{\lambda}_{\rho^{\prime},\sigma^{\prime},\tau^{\prime}}\left(s^{\prime}_{1}-\frac{\tau^{\prime}\gamma}{\rho^{\prime}\alpha}q_{2}-p_{1},s^{\prime}_{2}-p_{2}\right)ds_{1}ds_{2}ds^{\prime}_{1}ds^{\prime}_{2}\right\}\right.\nonumber\\
&&\left.\times d\theta d\phi d\psi dq_{1} dq_{2} dp_{1} dp_{2}\right]\kappa(\rho,\sigma,\tau)\kappa(\rho^{\prime},\sigma^{\prime},\tau^{\prime})d\rho d\sigma d\tau d\rho^{\prime}d\sigma^{\prime} d\tau^{\prime}\nonumber\\
=\lefteqn{\frac{(2\pi)^{3}}{N^2}\int_{(\rho,\sigma,\tau)}\int_{(\rho^{\prime},\sigma^{\prime},\tau^{\prime})}\left[\int_{\mathbb{R}^{4}}\left\{\int_{(s_{1},s_{2})\in\mathbb{R}^{2}}\int_{(s_{1}^{\prime},s_{2}^{\prime})\in\mathbb{R}^{2}}\delta(\rho-\rho^{\prime})\delta(\sigma-\sigma^{\prime})\delta(\tau-\tau^{\prime})\right.\right.}\nonumber\\
&&\left.\left.\times e^{i\alpha q_{1}(\rho s_{1}-\rho^{\prime}s_{1}^{\prime})+i\alpha q_{2}(\rho s_{2}-\rho^{\prime}s_{2}^{\prime})+i\beta p_{1}(\sigma s_{2}-\sigma^{\prime}s_{2}^{\prime})-\frac{i\alpha}{2}(\rho-\rho^{\prime})q_{1}p_{1}-\frac{i\alpha}{2}(\rho-\rho^{\prime})q_{2}p_{2}}\right.\right.\nonumber\\
&&\left.\left.\times e^{-\frac{i\gamma}{2}(\tau-\tau^{\prime})q_{1}q_{2}-\frac{i\beta}{2}(\sigma-\sigma^{\prime})p_{1}p_{2}}\overline{\hat{\lambda}_{\rho,\sigma,\tau}\left(s_{1}-\frac{\tau\gamma}{\rho\alpha}q_{2}-p_{1},s_{2}-p_{2}\right)}\hat{\chi}_{\rho,\sigma,\tau}(s_{1},s_{2})\right.\right.\nonumber\\
&&\left.\left.\times\overline{\hat{\chi}_{\rho^{\prime},\sigma^{\prime},\tau^{\prime}}(s^{\prime}_{1},s^{\prime}_{2})}\hat{\lambda}_{\rho^{\prime},\sigma^{\prime},\tau^{\prime}}\left(s^{\prime}_{1}-\frac{\tau^{\prime}\gamma}{\rho^{\prime}\alpha}q_{2}-p_{1},s^{\prime}_{2}-p_{2}\right)ds_{1}ds_{2}ds^{\prime}_{1}ds^{\prime}_{2}\right\}\right.\nonumber\\
&&\left.\times dq_{1} dq_{2} dp_{1} dp_{2}\right]\kappa(\rho,\sigma,\tau)\kappa(\rho^{\prime},\sigma^{\prime},\tau^{\prime})d\rho d\sigma d\tau d\rho^{\prime}d\sigma^{\prime} d\tau^{\prime}\nonumber\\
=\lefteqn{\frac{(2\pi)^{3}}{N^2}\int_{(\rho,\sigma,\tau)}\left[\int_{\mathbb{R}^{4}}\left\{\int_{(s_{1},s_{2})\in\mathbb{R}^{2}}\int_{(s_{1}^{\prime},s_{2}^{\prime})\in\mathbb{R}^{2}}e^{i\alpha\rho q_{1}(s_{1}-s_{1}^{\prime})+i\alpha\rho q_{2}(s_{2}-s_{2}^{\prime})+i\sigma\beta p_{1}(s_{2}-s_{2}^{\prime})}\right.\right.}\nonumber\\
&&\left.\left.\times\overline{\hat{\lambda}_{\rho,\sigma,\tau}\left(s_{1}-\frac{\tau\gamma}{\rho\alpha}q_{2}-p_{1},s_{2}-p_{2}\right)}\hat{\chi}_{\rho,\sigma,\tau}(s_{1},s_{2})\overline{\hat{\chi}_{\rho,\sigma,\tau}(s^{\prime}_{1},s^{\prime}_{2})}\right.\right.\nonumber\\
&&\left.\left.\times\hat{\lambda}_{\rho,\sigma,\tau}\left(s^{\prime}_{1}-\frac{\tau\gamma}{\rho\alpha}q_{2}-p_{1},s^{\prime}_{2}-p_{2}\right)ds_{1}ds_{2}ds^{\prime}_{1}ds^{\prime}_{2}\right\}dq_{1} dq_{2} dp_{1} dp_{2}\right]\nonumber\\
&&\times [\kappa(\rho,\sigma,\tau)]^{2}d\rho d\sigma d\tau\nonumber\\
=\lefteqn{\frac{(2\pi)^{4}}{N^2}\int_{(\rho,\sigma,\tau)}\frac{1}{|\rho|}\left[\int_{\mathbb{R}^{3}}\left\{\int_{(s_{1},s_{2})\in\mathbb{R}^{2}}\int_{(s_{1}^{\prime},s_{2}^{\prime})\in\mathbb{R}^{2}}\delta(s_{1}-s^{\prime}_{1})e^{i\alpha\rho q_{2}(s_{2}-s_{2}^{\prime})+i\sigma\beta p_{1}(s_{2}-s_{2}^{\prime})}\right.\right.}\nonumber\\
&&\left.\left.\times\overline{\hat{\lambda}_{\rho,\sigma,\tau}\left(s_{1}-\frac{\tau\gamma}{\rho\alpha}q_{2}-p_{1},s_{2}-p_{2}\right)}\hat{\chi}_{\rho,\sigma,\tau}(s_{1},s_{2})\overline{\hat{\chi}_{\rho,\sigma,\tau}(s^{\prime}_{1},s^{\prime}_{2})}\right.\right.\nonumber\\
&&\left.\left.\times\hat{\lambda}_{\rho,\sigma,\tau}\left(s^{\prime}_{1}-\frac{\tau\gamma}{\rho\alpha}q_{2}-p_{1},s^{\prime}_{2}-p_{2}\right)ds_{1}ds_{2}ds^{\prime}_{1}ds^{\prime}_{2}\right\}dq_{2} dp_{1} dp_{2}\right]\nonumber\\
&&\times [\kappa(\rho,\sigma,\tau)]^{2}d\rho d\sigma d\tau\nonumber\\
=\lefteqn{\frac{(2\pi)^{4}}{N^2}\int_{(\rho,\sigma,\tau)}\frac{1}{|\rho|}\left[\int_{\mathbb{R}^{3}}\left\{\int_{(s_{1},s_{2})\in\mathbb{R}^{2}}\int_{s_{2}^{\prime}\in\mathbb{R}}e^{i\alpha\rho q_{2}(s_{2}-s_{2}^{\prime})+i\sigma\beta p_{1}(s_{2}-s_{2}^{\prime})}\right.\right.}\nonumber\\
&&\left.\left.\times\overline{\hat{\lambda}_{\rho,\sigma,\tau}\left(s_{1}-\frac{\tau\gamma}{\rho\alpha}q_{2}-p_{1},s_{2}-p_{2}\right)}\hat{\chi}_{\rho,\sigma,\tau}(s_{1},s_{2})\overline{\hat{\chi}_{\rho,\sigma,\tau}(s_{1},s^{\prime}_{2})}\right.\right.\nonumber\\
&&\left.\left.\times\hat{\lambda}_{\rho,\sigma,\tau}\left(s_{1}-\frac{\tau\gamma}{\rho\alpha}q_{2}-p_{1},s^{\prime}_{2}-p_{2}\right)ds_{1}ds_{2}ds^{\prime}_{2}\right\}dq_{2} dp_{1} dp_{2}\right]\nonumber\\
&&\times [\kappa(\rho,\sigma,\tau)]^{2}d\rho d\sigma d\tau.
\end{eqnarray}
Introducing the following change of variable
\begin{equation}\label{eq:change-var-planchrl}
p_{1}+\frac{\tau\gamma}{\rho\alpha}q_{2}=\Pi_{1},
\end{equation}
(\ref{eq:planchrl-dervtn}) now reduces to
\begin{eqnarray}\label{eq:plnchrl-dervtn-last-part}
\lefteqn{\frac{(2\pi)^{4}}{N^2}\int_{(\rho,\sigma,\tau)}\frac{1}{|\rho|}\left[\int_{(q_{2},\Pi_{1},p_{2})\in\mathbb{R}^{3}}\left\{\int_{(s_{1},s_{2})\in\mathbb{R}^{2}}\int_{s_{2}^{\prime}\in\mathbb{R}}e^{i\sigma\beta\Pi_{1}(s_{2}-s_{2}^{\prime})}\right.\right.}\nonumber\\
&&\left.\left.\times e^{i\left(\rho\alpha-\frac{\sigma\beta\gamma\tau}{\rho\alpha}\right) q_{2}(s_{2}-s_{2}^{\prime})}\overline{\hat{\lambda}_{\rho,\sigma,\tau}\left(s_{1}-\Pi_{1},s_{2}-p_{2}\right)}\hat{\chi}_{\rho,\sigma,\tau}(s_{1},s_{2})\overline{\hat{\chi}_{\rho,\sigma,\tau}(s_{1},s^{\prime}_{2})}\right.\right.\nonumber\\
&&\left.\left.\times\hat{\lambda}_{\rho,\sigma,\tau}\left(s_{1}-\Pi_{1},s^{\prime}_{2}-p_{2}\right)ds_{1}ds_{2}ds^{\prime}_{2}\right\}dq_{2} d\Pi_{1} dp_{2}\right][\kappa(\rho,\sigma,\tau)]^{2}d\rho d\sigma d\tau\nonumber\\
=\lefteqn{\frac{(2\pi)^{5}}{N^2}\int_{(\rho,\sigma,\tau)}\frac{\alpha^{2}}{|\rho^{2}\alpha^{2}-\sigma\beta\gamma\tau|}\left[\int_{(\Pi_{1},p_{2})\mathbb{R}^{2}}\left\{\int_{(s_{1},s_{2})\in\mathbb{R}^{2}}\int_{s_{2}^{\prime}\in\mathbb{R}}\delta(s_{2}-s^{\prime}_{2})e^{i\sigma\beta\Pi_{1}(s_{2}-s_{2}^{\prime})}\right.\right.}\nonumber\\
&&\left.\left.\times \overline{\hat{\lambda}_{\rho,\sigma,\tau}\left(s_{1}-\Pi_{1},s_{2}-p_{2}\right)}\hat{\chi}_{\rho,\sigma,\tau}(s_{1},s_{2})\overline{\hat{\chi}_{\rho,\sigma,\tau}(s_{1},s^{\prime}_{2})}\right.\right.\nonumber\\
&&\left.\left.\times\hat{\lambda}_{\rho,\sigma,\tau}\left(s_{1}-\Pi_{1},s^{\prime}_{2}-p_{2}\right)ds_{1}ds_{2}ds^{\prime}_{2}\right\}d\Pi_{1} dp_{2}\right][\kappa(\rho,\sigma,\tau)]^{2}d\rho d\sigma d\tau\nonumber\\
=\lefteqn{\frac{(2\pi)^{5}}{N^2}\int_{(\rho,\sigma,\tau)}\frac{\alpha^{2}}{|\rho^{2}\alpha^{2}-\sigma\beta\gamma\tau|}\left[\int_{(\Pi_{1},p_{2})\mathbb{R}^{2}}\left\{\int_{(s_{1},s_{2})\in\mathbb{R}^{2}}\overline{\hat{\lambda}_{\rho,\sigma,\tau}\left(s_{1}-\Pi_{1},s_{2}-p_{2}\right)}\right.\right.}\nonumber\\
&&\left.\left.\times\hat{\chi}_{\rho,\sigma,\tau}(s_{1},s_{2})\overline{\hat{\chi}_{\rho,\sigma,\tau}(s_{1},s_{2})}
\hat{\lambda}_{\rho,\sigma,\tau}\left(s_{1}-\Pi_{1},s_{2}-p_{2}\right)ds_{1}ds_{2}\right\}d\Pi_{1} dp_{2}\right]\nonumber\\
&&\times[\kappa(\rho,\sigma,\tau)]^{2}d\rho d\sigma d\tau\nonumber\\
=\lefteqn{\frac{(2\pi)^{5}}{N^{2}}\int_{(\rho,\sigma,\tau)\in \mathbb{R}^{*}\times\mathbb{R}^{*}\times\mathbb{R}^{*}}\frac{\alpha^{2}[\kappa(\rho,\sigma,\tau)]^{2}}{|\rho^{2}\alpha^{2}-\sigma\beta\gamma\tau|}\|\hat{\lambda}_{\rho,\sigma,\tau}\|^{2}\|\hat{\chi}_{\rho,\sigma,\tau}\|^{2}d\rho d\sigma d\tau.}
\end{eqnarray}
Now comparing (\ref{eq:plnchrl-dervtn-last-part}) with the right side of (\ref{eq:ortho-con-presnt-secnario}), one immediately obtains the following
\begin{eqnarray*}\label{eqn:planchrl-DM-final-eqn}
\kappa&=&\frac{|\rho^{2}\alpha^{2}-\gamma\beta\sigma\tau|}{\alpha^{2}}\\
N&=& (2\pi)^{\frac{5}{2}}.
\end{eqnarray*}
\qed

\bigskip
\prf {\bf of Theorem \ref{Theorem:NC-Wig-func}
}

We choose an arbitrary  element $g$ of the group $\g$ as $(-\theta,-\phi,-\psi,-\bq,-\bp)$ so that the inverse group element $g^{-1}$ is given by $(\theta,\phi,\psi,\bq,\bp)$. Now, using the definition given in (\ref{wig-fcn-def2}), the Wigner function of $\g$ restricted to the 4-dimensional coadjoint orbits $\mathcal{O}^{k_{1},k_{2},k_{3}}_{4}$ reads

\begin{align}
\MoveEqLeft W(|\hat{\chi}_{k_1,k_2,k_3}\rangle\langle\hat{\lambda}_{k_1,k_2,k_3}|;k_{1}^{*},k_{2}^{*},k_{3}^{*},k_{4}^{*};k_1,k_2,k_3)\nonumber\\
&=\frac{\vert\alpha\vert}{(2\pi)^{6}\vert k_{1}^{2}\alpha^{2}-k_{2}k_{3}\gamma\beta\vert^{\frac{1}{2}}}\int_{\mathbb{R}^{7}}e^{-i\alpha(-k_{1}^{*}p_{1}-k_{2}^{*}p_{2}-k_{3}^{*}q_{1}-k_{4}^{*}q_{2})}e^{-i(-k_{1}\theta-k_{2}\phi-k_{3}\psi)}\nonumber\\
&\times\left[\int_{(\omega,\nu,\mu)}\int_{(s_{1},s_{2})\in \mathbb{R}^{2}}\!\!\!\!e^{-i\omega(\theta+\alpha q_{1}s_{1}+\alpha q_{2}s_{2}-\frac{\alpha}{2}q_{1}p_{1}-\frac{\alpha}{2}q_{2}p_{2})}e^{-i\nu(\phi+\beta p_{1}s_{2}-\frac{\beta}{2}p_{1}p_{2})}e^{-i\mu(\psi-\frac{\gamma}{2}q_{1}q_{2})}\right.\nonumber\\
&\left.\times\overline{\hat{\lambda}_{\omega,\nu,\mu}(s_{1},s_{2})}\hat{\chi}_{\omega,\nu,\mu}\left(s_{1}-p_{1}-\frac{\mu\gamma}{\omega\alpha}q_{2},s_{2}-p_{2}\right)\frac{\vert\omega^{2}\alpha^{2}-\gamma\beta\nu\mu\vert}{\alpha^{2}}
ds_{1}ds_{2}d\omega d\nu d\mu\right]\nonumber\\
&\times d\theta d\phi d\psi dq_{1}dq_{2}dp_{1}dp_{2}\nonumber\\
&=\frac{(2\pi)^{3}\vert\alpha\vert}{(2\pi)^{6}\vert k_{1}^{2}\alpha^{2}-k_{2}k_{3}\gamma\beta\vert^{\frac{1}{2}}}\int_{\mathbb{R}^{4}}e^{ i\alpha(k_{1}^{*}p_{1}+k_{2}^{*}p_{2}+k_{3}^{*}q_{1}+k_{4}^{*}q_{2})}\delta(\omega-k_{1})\delta(\nu-k_{2})\delta(\mu-k_{3})\nonumber\\
&\times\left[\int_{(\omega,\nu,\mu)}\int_{(s_{1},s_{2})\in \mathbb{R}^{2}}e^{-i\omega(\alpha q_{1}s_{1}+\alpha q_{2}s_{2}-\frac{\alpha}{2}q_{1}p_{1}-\frac{\alpha}{2}q_{2}p_{2})}e^{-i\nu(\beta p_{1}s_{2}-\frac{\beta}{2}p_{1}p_{2})}e^{\frac{i\mu\gamma}{2}q_{1}q_{2}}\right.\nonumber\\
&\left.\times\overline{\hat{\lambda}_{\omega,\nu,\mu}(s_{1},s_{2})}\hat{\chi}_{\omega,\nu,\mu}\left(s_{1}-p_{1}-\frac{\mu\gamma}{\omega\alpha}q_{2},s_{2}-p_{2}\right)\frac{\vert\omega^{2}\alpha^{2}-\gamma\beta\nu\mu\vert}{\alpha^{2}}
ds_{1}ds_{2}d\omega d\nu d\mu\right]\nonumber\\
&\times dq_{1}dq_{2}dp_{1}dp_{2}\nonumber\\
&=\frac{\vert k_{1}^{2}\alpha^{2}-k_{2}k_{3}\gamma\beta\vert^{\frac{1}{2}}}{(2\pi)^{3}\vert\alpha\vert}\int_{\mathbb{R}^{4}}e^{i\alpha(k_{1}^{*}p_{1}+k_{2}^{*}p_{2}+k_{3}^{*}q_{1}+k_{4}^{*}q_{2})}\left[\int_{\mathbb{R}^{2}}e^{-ik_{1}\alpha(q_{1}s_{1}+q_{2}s_{2}-\frac{1}{2}q_{1}p_{1}-\frac{1}{2}q_{2}p_{2})}\right.\nonumber\\
&\left.\times e^{-ik_{2}\beta(p_{1}s_{2}-\frac{1}{2}p_{1}p_{2})}e^{\frac{ik_{3}\gamma}{2}q_{1}q_{2}}\overline{\hat{\lambda}_{k_{1},k_{2},k_{3}}(s_{1},s_{2})}\right.\nonumber\\
&\left.\times\hat{\chi}_{k_{1},k_{2},k_{3}}\left(s_{1}-p_{1}-\frac{k_{3}\gamma}{k_{1}\alpha}q_{2},s_{2}-p_{2}\right)
ds_{1}ds_{2}\right]dq_{1}dq_{2}dp_{1}dp_{2}\nonumber\\
&=\frac{|k_{1}^{2}\alpha^{2}-k_{2}k_{3}\gamma\beta|^{\frac{1}{2}}}{(2\pi)^{3}\vert\alpha\vert}\int_{\mathbb{R}^{4}}e^{ i\alpha(k_{1}^{*}p_{1}+k_{2}^{*}p_{2}+k_{3}^{*}q_{1}+k_{4}^{*}q_{2})}\left[\int_{\mathbb{R}^{2}}e^{-ik_{1}\alpha q_{1}s_{1}}\right.\nonumber\\
&\left.\times e^{-ik_{1}\alpha\left(q_{2}+\frac{k_{2}\beta}{k_{1}\alpha}p_{1}\right)s_{2}+\frac{ik_{1}\alpha}{2}\left(q_{2}+\frac{k_{2}\beta}{k_{1}\alpha}p_{1}\right)p_{2}+\frac{ik_{1}\alpha}{2}\left(p_{1}+\frac{k_{3}\gamma}{k_{1}\alpha}q_{2}\right)q_{1}}\overline{\hat{\lambda}_{k_{1},k_{2},k_{3}}(s_{1},s_{2})}\right.\nonumber\\
&\left.\times\hat{\chi}_{k_{1},k_{2},k_{3}}\left(s_{1}-p_{1}-\frac{k_{3}\gamma}{k_{1}\alpha}q_{2},s_{2}-p_{2}\right)
ds_{1}ds_{2}\right]dq_{1}dq_{2}dp_{1}dp_{2}.
\label{eq:NC-Wig-func-dervtn-first-prt}
\end{align}
Introduce the following change of variables
\begin{equation}\label{eq:change-of-var-NC-Wig-func}
\begin{aligned}
\widetilde{q}_{1}&=q_{1}\\
\widetilde{q}_{2}&=q_{2}+\frac{k_{2}\beta}{k_{1}\alpha}p_{1}\\
\widetilde{p}_{1}&=p_{1}+\frac{k_{3}\gamma}{k_{1}\alpha}q_{2}\\
\widetilde{p}_{2}&=p_{2},
\end{aligned}
\end{equation}
so that the associated change in measure is given by
\begin{equation}\label{measure-change-NC-Wig-func}
dq_{1}\;dq_{2}\;dp_{1}\;dp_{2}=\frac{k_{1}^{2}\alpha^{2}}{|k_{1}^{2}\alpha^{2}-k_{2}k_{3}\beta\gamma|}d\widetilde{q}_{1}\;d\widetilde{q}_{2}\;d\widetilde{p}_{1}\;d\widetilde{p}_{2}.
\end{equation}

Also, denote $\Lambda$, for the sake of convenience, as
\begin{equation}\label{notation-factor}
\Lambda=\frac{\vert\alpha\vert}{\vert k_{1}\alpha^{2}-k_{2}k_{3}\gamma\beta\vert^{\frac{1}{2}}}
\end{equation}

Now,using (\ref{eq:change-of-var-NC-Wig-func}) and (\ref{measure-change-NC-Wig-func}) in (\ref{eq:NC-Wig-func-dervtn-first-prt}), one immediately finds that
\begin{align}\label{NC-Wig-funct-last-part}
\MoveEqLeft W(|\hat{\chi}_{\rho,\sigma,\tau}\rangle\langle\hat{\lambda}_{\rho,\sigma,\tau}|;k_{1}^{*},k_{2}^{*},k_{3}^{*},k_{4}^{*};k_1,k_2,k_3)\nonumber\\
&=\frac{k_{1}^{2}\Lambda}{(2\pi)^{3}}\int_{\mathbb{R}^{4}}e^{ i\alpha k_{3}^{*}\widetilde{q}_{1}+i\alpha k_{2}^{*}\widetilde{p}_{2}+\frac{ik_{1}^{*}k_{1}^{2}\alpha^{3}}{k_{1}^{2}\alpha^{2}-k_{2}k_{3}\beta\gamma}\left(\widetilde{p}_{1}-\frac{k_{3}\gamma}{k_{1}\alpha}\widetilde{q}_{2}\right)}\nonumber\\
&\times e^{\frac{i k_{4}^{*}k_{1}^{2}\alpha^{3}}{k_{1}^{2}\alpha^{2}-k_{2}k_{3}\beta\gamma}\left(\widetilde{q}_{2}-\frac{k_{2}\beta}{k_{1}\alpha}\widetilde{p}_{1}\right)+\frac{ik_{1}\alpha}{2}(\widetilde{q}_{1}\widetilde{p}_{1}+\widetilde{q}_{2}\widetilde{p}_{2})}\left[\int_{\mathbb{R}^{2}}e^{-ik_{1}\alpha (\widetilde{q}_{1}s_{1}+\widetilde{q}_{2}s_{2})} \overline{\hat{\lambda}_{k_{1},k_{2},k_{3}}(s_{1},s_{2})}\right.\nonumber\\
&\left.\times\hat{\chi}_{k_{1},k_{2},k_{3}}\left(s_{1}-\widetilde{p}_{1},s_{2}-\widetilde{p}_{2}\right)
ds_{1}ds_{2}\right]d\widetilde{q}_{1}\;d\widetilde{q}_{2}\;d\widetilde{p}_{1}\;d\widetilde{p}_{2}\nonumber\\
&=\frac{k_{1}^{2}\Lambda}{(2\pi)^{3}}\int_{\mathbb{R}^{2}}\left[\int_{\mathbb{R}^{4}}e^{i\alpha\left(\frac{k_{1}^{*}k_{1}^{2}\alpha^{2}-k_{4}^{*}k_{1}k_{2}\alpha\beta}{k_{1}^{2}\alpha^{2}-k_{2}k_{3}\beta\gamma}\right)\widetilde{p}_{1}+i\alpha k_{2}^{*}\widetilde{p}_{2}-i\alpha k_{1}\left(s_{2}-\frac{1}{2}\widetilde{p}_{2}-\frac{k_{1}k_{4}^{*}\alpha^{2}-k_{1}^{*}k_{3}\alpha\gamma}{k_{1}^{2}\alpha^{2}-k_{2}k_{3}\beta\gamma}\right)\widetilde{q}_{2}}\right.\nonumber\\
&\left.\times e^{-i\alpha k_{1}\left(s_{1}-\frac{1}{2}\widetilde{p}_{1}-\frac{k_{3}^{*}}{k_{1}}\right)\widetilde{q}_{1}}\overline{\hat{\lambda}_{k_{1},k_{2},k_{3}}(s_{1},s_{2})}\hat{\chi}_{k_{1},k_{2},k_{3}}(s_{1}-\widetilde{p}_{1},s_{2}-\widetilde{p}_{2})\right.\nonumber\\
&\left.\times d\widetilde{p}_{1}\;d\widetilde{p}_{2}\;d\widetilde{q}_{1}\;d\widetilde{q}_{2}\right]ds_{1}ds_{2}\nonumber\\
&=\frac{k_{1}^{2}\Lambda}{2\pi}\int_{\mathbb{R}^{2}}\left[\int_{\mathbb{R}^{2}}e^{i\alpha\left(\frac{k_{1}^{*}k_{1}^{2}\alpha^{2}-k_{4}^{*}k_{1}k_{2}\alpha\beta}{k_{1}^{2}\alpha^{2}-k_{2}k_{3}\beta\gamma}\right)\widetilde{p}_{1}+i\alpha k_{2}^{*}\widetilde{p}_{2}}\frac{1}{|k_{1}|}\delta\left(s_{1}-\frac{1}{2}\widetilde{p}_{1}-\frac{k_{3}^{*}}{k_{1}}\right)\right.\nonumber\\
&\left.\times\frac{1}{|k_{1}|}\delta\left(s_{2}-\frac{1}{2}\widetilde{p}_{2}-\frac{k_{1}k_{4}^{*}\alpha^{2}-k_{1}^{*}k_{3}\alpha\gamma}{k_{1}^{2}\alpha^{2}-k_{2}k_{3}\beta\gamma}\right)\overline{\hat{\lambda}_{k_{1},k_{2},k_{3}}(s_{1},s_{2})}\right.\nonumber\\
&\left.\times\hat{\chi}_{k_{1},k_{2},k_{3}}(s_{1}-\widetilde{p}_{1},s_{2}-\widetilde{p}_{2})d\widetilde{p}_{1}\;d\widetilde{p}_{2}\right]ds_{1}ds_{2}\nonumber\\
&=\frac{\Lambda}{2\pi}\int_{\mathbb{R}^{2}}e^{i\alpha\left(\frac{k_{1}^{*}k_{1}^{2}\alpha^{2}-k_{4}^{*}k_{1}k_{2}\alpha\beta}{k_{1}^{2}\alpha^{2}-k_{2}k_{3}\beta\gamma}\right)\widetilde{p}_{1}+i\alpha k_{2}^{*}\widetilde{p}_{2}}\nonumber\\
&\times\overline{\hat{\lambda}_{k_{1},k_{2},k_{3}}\left(\frac{1}{2}\widetilde{p}_{1}+\frac{k_{3}^{*}}{k_{1}},\frac{1}{2}\widetilde{p}_{2}+\frac{k_{1}k_{4}^{*}\alpha^{2}-k_{1}^{*}k_{3}\alpha\gamma}{k_{1}^{2}\alpha^{2}-k_{2}k_{3}\beta\gamma}\right)}\nonumber\\
&\times\hat{\chi}_{k_{1},k_{2},k_{3}}\left(-\frac{1}{2}\widetilde{p}_{1}+\frac{k_{3}^{*}}{k_{1}},-\frac{1}{2}\widetilde{p}_{2}+\frac{k_{1}k_{4}^{*}\alpha^{2}-k_{1}^{*}k_{3}\alpha\gamma}{k_{1}^{2}\alpha^{2}-k_{2}k_{3}\beta\gamma}\right)d\widetilde{p}_{1}\;d\widetilde{p}_{2}.\nonumber
\end{align}
Substituting $\Lambda$ from (\ref{notation-factor}) yields the desired result.
\qed
\section*{Acknowledgements}
One of the authors (SHHC) would like to acknowledge fruitful discussions with Hartmut F\"{u}hr.

%\bibliographystyle{plain}
%\bibliography{biblia}

\begin{thebibliography}{99}


\bibitem{AlAtChWo}
S.T. Ali, N.M. Atakishiyev, S.M. Chumakov and K.B. Wolf
\newblock {\em The Wigner
  function for general Lie groups and the wavelet transform\/,}
\newblock {Ann. H. Poincar\'e}  {\bf 1}, 685-714 (2000).


\bibitem{AlKrMu}
S.T.Ali, A.E. Krasowska and R. Murenzi
\newblock {\em Wigner functions from the
  two-dimensional wavelets group\/,}
\newblock {J. Opt. Soc. Am.},  {\bf A17},  1-11 (2000).


\bibitem{plancherel}
S. T. Ali, H. F\"{u}hr, and A. E. Krasowska,
\newblock {\em Plancherel inversion as unified approach to Wavelet transforms and Wigner functions},
\newblock {Ann. Inst. Henri Poincar\'e}, {\bf 4}, 1015--1050 (2003).



\bibitem{bastosjmp}
C. Bastos, O. Bertolami, N.C. Dias, J.N. Prata,
\newblock {\em Weyl-Wigner Formulation of Noncommutative Quantum
Mechanics.}
\newblock {J. Math. Phys.}, {\bf 49}, 072101 (2008).



\bibitem{bastoscmp}
 C. Bastos, N.C. Dias, and J.N. Prata,
\newblock {\em Wigner measures in noncommutative quantum
mechanics.}
\newblock {Comm. Math. Phys.}, {\bf 299}, 709--740 (2010).


\bibitem{ncqmjmp}
S. H. H. Chowdhury and S. T. Ali,
\newblock {\em The symmetry groups of noncommutative quantum mechanics and coherent state quantization.}
\newblock {J. Math. Phys.}, {\bf 54}, 032101 (21pp) (2013).



\bibitem{ncqmjpa}
S. H. H. Chowdhury and S. T. Ali,
\newblock {\em Triply Extended Group of Translations of $\mathbb{R}^{4}$ as Defining Group of NCQM: relation to various gauges.}
\newblock J. Phys. A: Math. Theor., {\bf 47}, 085301 (29pp) (2014).


\bibitem{nctori}
S. H. H. Chowdhury,
\newblock {\em On the plethora of representations arising in noncommutative quantum mechanics and an explicit construction of noncommutative 4-tori.}
\newblock To appear.




\bibitem{delducetalt}
F. Delduc, Q. Duret, F. Gieres and M. Lafran\c{c}ois,
\newblock {\em Magnetic fields in noncommutative quantum mechanics\/,}
\newblock {J. Phys. Conf. Ser.} {\bf 103}, 012020 (2008).

\bibitem{Du-Mo}
M. Duflo and C.C. Moore,
\newblock{\em On the regular representation of a nonunimodular locally compact group\/,}
\newblock J. Funct. Anal. {\bf 21},  209--243  (1976).

\bibitem{Fuhrbook}
H. F\"{u}hr,
\newblock {\em Abstract Harmonic Analysis of Continuous Wavelet Transforms},
\newblock {Springer Lecture Notes in Mathematics.} {\bf 1863}, Springer Verlag, Heidelberg, 2005.


\bibitem{wigfuncchinese}
S.C. Jing, T.H. Heng  and  F. Zuo,
\newblock {\em A new form of Wigner functions on the noncommutative space\/,}
\newblock {Phys. Lett. A} {\bf 335}, 185--190 (2005).


\bibitem{Kirillovbook}
A.A. Kirillov,
\newblock {\em Lectures on the Orbit Method},
\newblock American Math. Soc., 2004.

\bibitem{scholtzt}
F.G Scholtz, L. Gouba., A. Hafver and C.M Rohwer,
\newblock {\em Formulation, interpretation and application of non-commutative
quantum mechanics\/,}
\newblock J. Phys. A: Math. Theor., {\bf 42}, 175303 (13pp)  (2009).

\bibitem{Wigner} E.P. Wigner,
\newblock {\em On the quantum correction for
      thermodynamic equilibrium\/,}
\newblock {Phys. Rev.}, {\bf 40}, 749--759   (1932) .
\end{thebibliography}

\end{document}